\documentclass[a4paper,11pt]{article}
\pdfoutput=1 

\usepackage{jheppub}

\usepackage{amsmath,amssymb,amsfonts, physics}
\usepackage{mathrsfs}
\usepackage{bbm}
\usepackage{slashed}
\usepackage{graphicx}
\usepackage{comment}
\usepackage{verbatim}
\usepackage[T1]{fontenc}
\usepackage[utf8]{inputenc}
\usepackage{mathbbol}
\usepackage[dvipsnames]{xcolor}
\usepackage[normalem]{ulem}
\usepackage{acronym}
\usepackage[nameinlink]{cleveref}

\graphicspath{{Figures/}}

\newcommand{\Ftwo}{\ensuremath{\mathcal F_2}}
\newcommand{\Ffour}{\ensuremath{\mathcal F_4}}
\newcommand{\regulatortensor}{\ensuremath{\mathfrak R}}
\newcommand{\regulatorfunction}{\ensuremath{\mathcal R}}
\newcommand{\eulerterm}{\ensuremath{\mathfrak E}}

\newcommand{\scaleparameter}{\ensuremath{c_l}}
\newcommand{\WC}[2]{\ensuremath{w_{#1}^{#2}}}

\newcommand{\eg}{e.g.}
\newcommand{\ie}{i.e.}

\newacro{UV}[UV]{ultraviolet}
\newacro{IR}[IR]{infrared}
\newacro{AS}[AS]{Asymptotic Safety}
\newacro{QG}[QG]{quantum gravity}
\newacro{QFT}[QFT]{quantum field theory}
\acrodefplural{QFT}{quantum field theories}
\newacro{EFT}[EFT]{effective field theory}
\acrodefplural{EFT}{effective field theories}
\newacro{GR}[GR]{General Relativity}
\newacro{FRG}[FRG]{Functional Renormalization Group}
\newacro{RG}[RG]{renormalization group}
\newacro{WGC}[WGC]{weak gravity conjecture}
\newacro{FP}[FP]{fixed point}
\newacro{GFP}[GFP]{Gaussian fixed point}
\newacro{MFP}[MFP]{matter fixed point}
\newacro{ST}[ST]{String Theory}

\title{\boldmath Unearthing the intersections: positivity bounds, weak gravity conjecture, and asymptotic safety landscapes from photon-graviton flows}

\author[a]{Benjamin Knorr\,\href{https://orcid.org/0000-0001-6700-6501}{\protect \includegraphics[scale=.07]{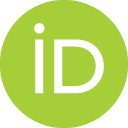}}\,}
\author[b]{and Alessia Platania\,\href{https://orcid.org/0000-0001-7789-344X}{\protect \includegraphics[scale=.07]{ORCIDiD_icon128x128.png}}\,}
\affiliation[a]{Nordita, Stockholm University and KTH Royal Institute of Technology, Hannes Alfv\'ens v\"ag 12, SE-106 91 Stockholm, Sweden}
\affiliation[b]{Niels Bohr International Academy, The Niels Bohr Institute, Blegdamsvej 17, DK-2100 Copenhagen \O, DENMARK}

\emailAdd{benjamin.knorr@su.se}
\emailAdd{alessia.platania@nbi.ku.dk}

\abstract{We compute the asymptotic safety landscape stemming from ultraviolet-com\-plete photon-graviton flows in a field theoretic setup, and we confront it with the weak gravity conjecture and, for the first time, with positivity bounds. At fourth order in derivatives, we find two gravitational fixed points providing viable ultraviolet completions for the theory. One of them comes with a single relevant direction, which sets the scale of quantum gravity. The corresponding sub-landscape is a single point. The second fixed point yields a richer sub-landscape of effective theories, most of which is described by an approximately straight line in the space of dimensionless Wilson coefficients. We additionally discover that: \textit{(i)} the two sub-landscapes are continuously connected via a small ``candy cane'' regime, and the whole asymptotic safety landscape falls onto a plane; this is consistent with earlier findings and could be a universal feature in Asymptotic Safety; \textit{(ii)} in such a field-theoretic setup, the Euler coupling plays a special role, as it is unconstrained by quantum scale invariance, but can enter off-shell bounds such as entropy-based positivity constraints; \textit{(iii)} Planck-scale-suppressed violations of both weak gravity and positivity bounds occur across the landscape. The latter result resonates with expectations grounded on effective field theory arguments.}

\makeatletter
\gdef\@fpheader{}
\makeatother
\begin{document} 
\maketitle
\flushbottom

\clearpage

\section{Introduction} \label{sec:intro}

One of the most challenging open problems in theoretical physics is to understand the gravitational interaction at the smallest distance scales. At Planckian scales, we expect gravity to shed off its classical behavior and display quantum properties, similarly to other gauge and matter fields. Different approaches to \ac{QG} attempt to describe the phenomena at these scales starting from a variety of fundamental ideas and based on seemingly different frameworks. Fundamental research in each of these theories entails the investigation of their \ac{UV} details and, ideally, the use of top-down strategies to derive predictions from scratch. \Ac{EFT}, on the other hand, is a powerful mathematical formalism to describe physical phenomena involving particles and fields within a given energy range --- usually below a cutoff scale where new physics becomes relevant. It serves as a pragmatic approach for modeling complex systems while incorporating the effects of higher energy degrees of freedom or features through systematic expansions. 

The challenge lies in bridging the gap between these two frameworks: \ac{QG}, which governs the behavior of spacetime and gravitational interactions at the Planck scale, and \ac{EFT}, which parameterizes gravity-matter systems at much lower energy scales. 

One way to connect \ac{QG} to \ac{EFT} is through decoupling~\cite{Appelquist:1974tg}. This involves constructing \acp{EFT} that capture the low-energy dynamics of gravity while incorporating the effects of quantum fluctuations at shorter distances. By integrating out the high-energy modes, one can derive \acp{EFT} that manifest as low-energy approximations to the underlying \ac{QG} theory. A paradigmatic example of this mechanism is realized within the \ac{AS} program for \ac{QG}~\cite{Bonanno:2020bil, Knorr:2022dsx, Eichhorn:2022gku, Morris:2022btf, Martini:2022sll, Wetterich:2022ncl, Platania:2023srt, Saueressig:2023irs, Pawlowski:2023gym,Bonanno:2024xne}, which builds on the framework of \ac{QFT}, and on the idea that gravity could be \ac{UV}-complete with respect to an interacting fixed point of the gravitational \ac{RG} flow. Within the \ac{AS} approach, much effort has been put into corroborating the existence of this fixed point~\cite{Reuter:1996cp, Falls:2014tra, Gies:2016con}, as well as in assessing its unitarity~\cite{Platania:2020knd,Bonanno:2021squ, Fehre:2021eob, Platania:2022gtt, Knorr:2023usb}, and its compatibility with matter~\cite{Dona:2013qba, Eichhorn:2017ylw, Eichhorn:2022jqj, Pastor-Gutierrez:2022nki}. Although \ac{RG} trajectories have been found that connect the fixed point with the \ac{GR} regime, and despite the natural embedding of the decoupling mechanism in \ac{AS}, a systematic study of the \ac{QG}-\ac{EFT} map in \ac{AS} is missing. Similarly, most of the other \ac{QG} approaches have devoted their main focus on understanding the \ac{UV} details of the theory, often in isolation from matter. 
\begin{figure}
    \centering
    \includegraphics[scale=0.8]{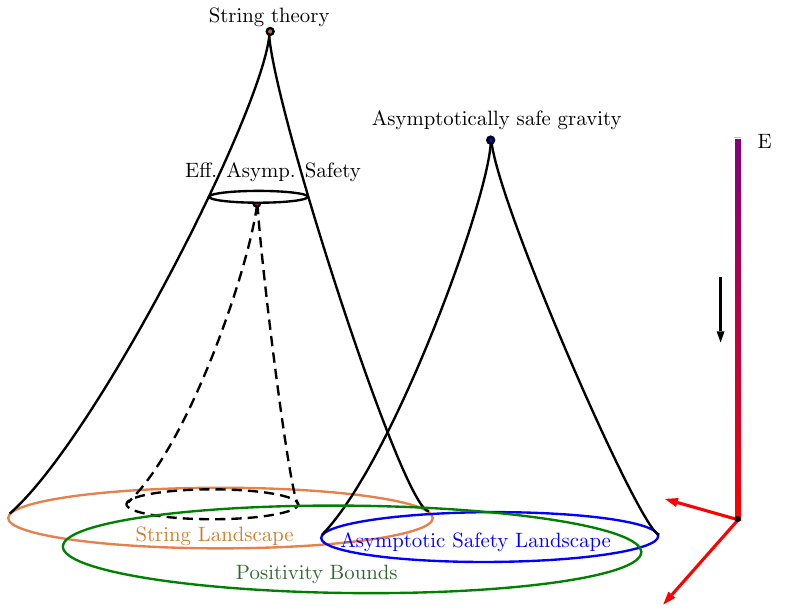}
    \caption{Sketch of the extensions of the swampland program~\cite{Vafa:2005ui} and the notion of landscape to other approaches to quantum gravity, in particular Asymptotically Safe Gravity. This paves the way to contrasting the asymptotic safety and string landscapes~\cite{Basile:2021krr} and assessing the possibility of Asymptotic Safety being a low-energy approximation of String Theory~\cite{Percacci:2010af,deAlwis:2019aud,Held:2020kze,Basile:2021euh,Basile:2021krk}, investigating the generality of swampland conjectures and string universality~\cite{Kumar:2009us,Montero:2020icj}, and testing Asymptotic Safety against important theoretical constraints such as positivity bounds~\cite{deRham:2022hpx}.}
    \label{fig:landscapes}
\end{figure}

The swampland program~\cite{Vafa:2005ui,Palti:2019pca} plays a crucial role in this endeavor, by providing constraints and guiding principles to construct \acp{EFT} that arise from consistent \ac{UV} completions of gravity. At its core, the swampland program seeks to identify the theoretical constraints --- formulated in terms of a set of conjectures --- that any consistent quantum theory of gravity must satisfy. Its name, ``swampland'', metaphorically reflects the idea that not all \acp{EFT} can arise from a consistent \ac{UV} theory. The program aims to delineate which theoretical frameworks belong to the swampland of inconsistent theories, and which ones originate from fundamental theories. The latter set identifies the ``landscape'' of consistent \acp{EFT}. Whether such a landscape solely contains consistent \acp{EFT} stemming from \ac{ST}, or more generally the set of \acp{EFT} generated by \textit{any} consistent \ac{QG} theory is an open question. According to the String Lamppost Principle~\cite{Kumar:2009us,Montero:2020icj}, all such \acp{EFT} coming from consistent \ac{QG} theories must have a stringy origin.

Although in the past years, the swampland program has sparked intense debate due to the use of conjectures, it has also inspired numerous research efforts. Its implications extend beyond the confines of \ac{ST}, influencing broader discussions in cosmology, particle physics, and beyond~\cite{Agrawal:2018own,Bedroya:2019tba,Angelantonj:2020pyr,Gonzalo:2021fma,Grana:2021zvf,Anchordoqui:2023wkm,Vafa:2024fpx}. In particular, despite the swampland program having emerged in the context of \ac{ST}, the general idea of selecting and constraining the set of \acp{EFT} compatible with and stemming from different \ac{UV} completions of gravity can~\cite{Basile:2021krr}, and should, be extended to other approaches to \ac{QG}. Deriving the landscapes from different approaches could indeed allow for \textit{(i)} a more efficient comparison with constraints derived from \ac{EFT}~\cite{deRham:2022hpx}, and \textit{(ii)} a more direct and clean dictionary to compare the predictions of different \ac{QG} approaches~\cite{Knorr:2021iwv} and frameworks, which in the \ac{UV} hardly talk to each other. 
Constructing the \ac{QG}-\ac{EFT} map and generalizing the notion of landscape to other \ac{QG} theories~\cite{Basile:2021krr} would additionally allow to investigate the validity of swampland conjectures beyond string models~\cite{deAlwis:2019aud,Basile:2021krr,Montero:2024sln}, and thus to test the String Lamppost Principle~\cite{Kumar:2009us,Montero:2020icj}. In a similar direction, analyzing the intersections between different \ac{QG} landscapes could reveal non-trivial connections between theories. For instance, if the \ac{AS} landscape would lie inside the string landscape, or if the two would have a non-trivial intersection~\cite{Basile:2021krr}, this could indicate that \ac{AS} be realized in the form of ``effective \ac{AS}''~\cite{Percacci:2010af,Held:2020kze}, \ie{}, as a low-energy approximation of \ac{ST}~\cite{deAlwis:2019aud,Basile:2021euh,Basile:2021krk}.
This overarching idea that generalizes the big picture of the swampland program is illustrated in \Cref{fig:landscapes}.

In this work, we take two key steps in the realization of this program. First, we extend the investigations of~\cite{Basile:2021krr} to a more sophisticated system and a full-fledged non-perturbative \ac{RG} computation. Second, we compare the \ac{AS} landscape not only with swampland conjectures --- specifically, the \ac{WGC}~\cite{Arkani-Hamed:2006emk,Montero:2018fns,Harlow:2022ich,Heidenreich:2024dmr} --- but also, for the first time, with positivity bounds~\cite{deRham:2022hpx}. Concretely, we will focus on a photon-graviton system at fourth order in a derivative expansion, while only including couplings that cannot be removed by a local field redefinition. Explicitly, the dynamics is encoded in the effective action
\begin{equation}\label{eq:action-intro}
\begin{aligned}
    \Gamma = &\int \text{d}^4x \, \sqrt{g} \, \Big[ \frac{R}{16\pi G_N} + G_{\eulerterm} \, \eulerterm - \Ftwo{} + G_{\Ftwo^2} \, (\Ftwo{})^2 + G_{\Ffour} \, \Ffour{} + G_{CFF} \, C^{\mu\nu\rho\sigma} F_{\mu\nu} F_{\rho\sigma}\Big] \, ,
\end{aligned}
\end{equation}
with
\begin{equation}\label{eq:operator-def}
    \Ftwo{} = \frac{1}{4} F^{\mu\nu} F_{\mu\nu} \, , \qquad \Ffour{} = \frac{1}{4} F^\mu_{\phantom{\mu}\nu} F^\nu_{\phantom{\nu}\rho} F^\rho_{\phantom{\rho}\sigma} F^\sigma_{\phantom{\sigma}\mu} \, , \qquad \eulerterm{} = R_{\mu\nu\rho\sigma}R^{\mu\nu\rho\sigma} - 4R_{\mu\nu}R^{\mu\nu} + R^2 \, .
\end{equation}
Here, $F_{\mu\nu}$ is the Abelian field strength tensor and $C_{\mu\nu\rho\sigma}$ is the Weyl tensor.
Along the lines of~\cite{Basile:2021krr}, we compute the beta functions and determine the \ac{AS} landscape as the \textit{hypersurface} of \acp{EFT} --- parameterized by the relevant dimensionless Wilson coefficients --- that are connected to \ac{UV}-complete \ac{RG} trajectories. We will then compare the Wilson coefficients in the landscape with standard positivity bounds~\cite{Bellazzini:2019xts, CarrilloGonzalez:2023cbf} and with a family of entropy-based positivity constraints~\cite{Cheung:2018cwt}, which contains the \ac{WGC} for black holes in the presence of higher derivatives~\cite{Kats:2006xp,Cheung:2018cwt,Hamada:2018dde} as a particular case.

Our system has two viable gravitational \ac{UV} fixed points, and hence the \ac{AS} landscape consists of two sub-landscapes. One fixed point comes with a single relevant direction, so that, once the scale of \ac{QG} is fixed, the resulting sub-landscape is a zero-parameter theory: a single point in the space of dimensionless Wilson coefficients. The second fixed point has two relevant directions, so that the corresponding sub-landscape is a line. In particular, this line is nearly straight. The straight approximation only ceases to work in a small region where the sub-landscape of the second fixed point bends to continuously connect to the single-point sub-landscape coming from the most predictive fixed point. Globally, the entire \ac{AS} landscape falls onto a plane --- a feature already observed in a previous work~\cite{Basile:2021krr}. Whether this is a coincidence or a universal feature of \ac{AS} is to be assessed by more extensive studies.

In agreement with general expectations from \ac{EFT}~\cite{Alberte:2020jsk,Herrero-Valea:2020wxz,Henriksson:2022oeu,Herrero-Valea:2022lfd}, we find that Planck-scale suppressed violations of weak gravity and positivity constraints can occur across the landscape. We will show that such violations are minimized by the most predictive sub-landscape. Finally, our work also highlights the special role played by the Euler coupling in \ac{AS}: while it is attached to a topological invariant and is thus generally unconstrained by the \ac{RG} flow~\cite{Falls:2020qhj, Knorr:2021slg}, it can enter off-shell positivity constraints such as those based on black hole entropy~\cite{Cheung:2018cwt}. Such positivity bounds can thus entail constraints on the Euler coupling, rather than tests of \ac{AS}. 

Our paper is organized as follows. In \Cref{sec:basics}, we discuss the concepts that are important to this work in more detail: landscapes, positivity bounds, and the weak gravity conjecture. \Cref{sec:landscape} contains our results on the fixed point structure, and an in-depth discussion of the ensuing landscapes. With these results in hand, in \Cref{sec:comparisonIR} and \Cref{sec:comparisonFlow} we confront our results with positivity, entropy, and weak gravity bounds. Finally, in \Cref{sec:conclusions} we summarize and discuss our results. The two appendices contain some more details about our technical setup and an analytical result.

\section{Landscapes, positivity bounds, and the weak gravity conjecture}\label{sec:basics}

This section introduces the basic concepts that we will need and use throughout the manuscript. This includes the notion of landscapes in \ac{QG} and an overview of the weak gravity and positivity bounds that we will intersect with the \ac{AS} landscape.

\subsection{Landscapes in String Theory and Asymptotic Safety}\label{sec:land-def}

The concept of landscape was first introduced in the context of \ac{ST}~\cite{Vafa:2005ui}, to indicate the set of \ac{IR} \acp{EFT} that admit a consistent \ac{UV} completion in the presence of gravity. The swampland is the complement of this set --- that is, the set of \acp{EFT} that cannot be \ac{UV}-completed when coupled to gravity. The difficulties in tracing the set of theories belonging to the string landscape via top-down derivations have led to the idea at the core of the swampland program~\cite{Palti:2019pca}: to determine the conditions or criteria that precisely select the \acp{EFT} belonging to the landscape. Such conditions are better known as ``swampland conjectures'', and stem from recurring patterns in stringy constructions, black hole physics, and the interrelations between different conjectures. In the following, we will assume that swampland conjectures precisely identify the string landscape.
\begin{figure}
    \centering
    \includegraphics[scale=0.8]{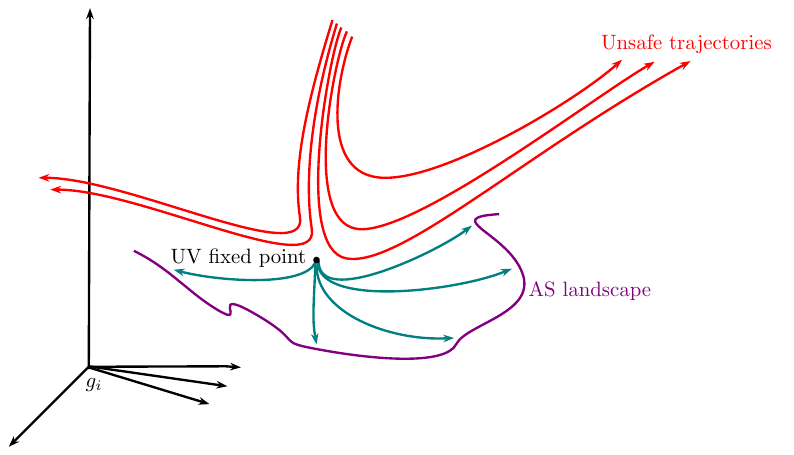}
    \caption{Generalization of the concept of landscape to \ac{AS}. In the space of dimensionless running couplings, the flow is controlled by one or more fixed points (black dot). Each fixed point with at least one \ac{UV}-attractive direction corresponds to a possible \ac{UV} completion. \ac{RG} trajectories whose \ac{UV} behavior is not governed by a fixed point (red lines) correspond to non-renormalizable theories. Asymptotically safe trajectories (petrol-green lines) are those starting at a fixed point in the \ac{UV}. The theory-space version of the \ac{AS} landscape stemming from a given fixed point (purple line) is the set of \ac{IR} endpoints of these trajectories. The corresponding \acp{EFT} in the space of dimensionless Wilson coefficients make up the \ac{AS} landscape of \Cref{fig:landscapes}.}
    \label{fig:AS-land}
\end{figure}

While the concept of landscape --- seen as the set of \acp{EFT} stemming from consistent gravity-matter \ac{UV} completions --- is universal, the strategy to determine it may depend on the specific approach to \ac{QG}. For a given approach to \ac{QG}, the goal is to predict the set of allowed Wilson coefficients in the corresponding low-energy \ac{EFT}. The first generalization of the concept of the \ac{QG} landscape beyond \ac{ST} has been put forth in~\cite{Basile:2021krr}, in the context of \ac{AS}~\cite{Reuter:2019byg}. As argued in~\cite{Basile:2021krr}, the \ac{AS} version of the string landscape is the set of Wilson coefficients in the effective action stemming from asymptotically safe \ac{RG} trajectories, cf. \Cref{fig:AS-land}. If more than one \ac{UV} fixed point exists, we will define the landscape as the union of the ``sub-landscapes'' identified by each viable \ac{UV} fixed point. These sub-landscapes are in general not contiguous, but can never overlap: they will typically be disjoint sets.

It is noteworthy that the concept of landscape in \ac{AS} is amenable to explicit computation. The \ac{FRG}~\cite{Dupuis:2020fhh} (see \Cref{sec:setup} for a brief review) is typically used in the field of \ac{AS} to corroborate the existence of interacting fixed points of the \ac{RG} flow, \ie{}, possible \ac{UV} completions of the theory. In a nutshell, the idea of the \ac{FRG} is to perform the path integral in the Wilsonian spirit by integrating out modes above an \ac{IR} cutoff scale $k$. All couplings thus acquire an \ac{RG}-scale dependence, and in the limit $k\to0$ --- corresponding to all fluctuations having been integrated out --- the standard effective action $\Gamma$ is obtained. As a consequence, the \ac{FRG} also provides a clear recipe to compute the Wilson coefficients belonging to the landscape~\cite{Basile:2021krr}: they are the \ac{IR} ($k\to0$) limits of the \ac{FRG}-running couplings $G_i(k)$, for all \ac{RG} trajectories departing from a given fixed point. Hence, \textit{computing the \ac{AS} landscape} consists of the following steps~\cite{Basile:2021krr}:
\begin{itemize}
	\item Solve the beta functions for \textit{all} couplings of the system for a sufficiently large number of (relevant) perturbations from the fixed point (or, initial conditions), uniformly covering the \ac{UV} critical surface.\footnote{See~\cite{Saueressig:2024ojx} for an alternative method.}
	\item Take the limit $k\to0$ of all these \ac{RG} trajectories (which are asymptotically safe by construction). If the initial conditions were appropriately selected, the resulting \ac{IR} points ought to cover the entire theory-space version of the \ac{AS} landscape, boundary to boundary. This theory-space landscape is the purple line of \Cref{fig:AS-land}.
	\item Switch to the space of \acp{EFT} parameterized by appropriate dimensionless Wilson coefficients, and determine the equations describing the hypersurface of asymptotically safe \acp{EFT} corresponding to the \ac{IR} points of the previous step. This is the \ac{AS} landscape of \Cref{fig:landscapes}, that can be directly confronted with the string landscape, positivity bounds, and observational constraints. 
\end{itemize}
If more than one \ac{UV} fixed point exists, the recipe above identifies only a sub-landscape, and thus it has to be iterated for each suitable fixed point. We shall return to this procedure in the next section, in the light of the definitions of positivity and entropy-based bounds.

\subsection{Positivity bounds}

Positivity bounds originally arose in the context of \ac{EFT}. The underlying idea is that consistency conditions for a fundamental theory, including locality, unitarity, analyticity, and Lorentz invariance, constrain the scattering amplitudes of the theory, even at scales above the cutoff where the \ac{EFT} is valid. More precisely, \emph{without} knowing the details of the \ac{UV} completion of the \ac{EFT}, one can still infer constraints on the \ac{IR} physics that follow from these consistency conditions. One strategy for deriving positivity bounds is to use the optical theorem together with the (assumed or known) branch cut and pole structure of scattering amplitudes in order to find contour integrals that are positive. This can then be mapped onto constraints on specific combinations of Wilson coefficients that are, by construction, gauge- and reparameterization-invariant. For a recent overview of the topic, see \eg{}~\cite{deRham:2022hpx}. For the theory that we consider in this work, the effective action is given by~\eqref{eq:action-intro}, and the relevant dimensionless Wilson coefficients are
\begin{equation}\label{eq:wilson-coeff-prime}
    \WC{\Ftwo^2}{} = \frac{G_{\Ftwo^2}}{(16\pi G_N)^2} \, , \quad \WC{\Ffour}{} = \frac{G_{\Ffour}}{(16\pi G_N)^2} \, , \quad \WC{C}{} = \frac{G_{CFF}}{16\pi G_N} \, ,
\end{equation}
Positivity bounds on these coefficients have been discussed \eg{} in~\cite{Bellazzini:2019xts, CarrilloGonzalez:2023cbf}.\footnote{For bounds of our system in arbitrary dimension, see~\cite{Bittar:2024xuc}.} In~\cite{CarrilloGonzalez:2023cbf}, bounds for the first two Wilson coefficients were derived, together with constraints for other operators that we do not consider in this work. With the \ac{EFT} cutoff~$\Lambda$, the dimensionless quantities
\begin{equation}
    f_2 = \frac{(16\pi G_N)^2\Lambda^4}{2}\,(\WC{\Ftwo^2}{}+\WC{\Ffour}{})  \, , \qquad g_2 =  \frac{(16\pi G_N)^2\Lambda^4}{2}\, (\WC{\Ftwo^2}{}+3\WC{\Ffour}{}) \, ,
\end{equation}
need to satisfy
\begin{equation}\label{eq:POS-CarrilloGonzalez}
    g_2 > |f_2| \, .
\end{equation}
On the other hand, in~\cite{Bellazzini:2019xts}, also the third Wilson coefficient has been included, and the scale has been chosen explicitly in terms of the Planck scale. 
The two bounds of~\cite{Bellazzini:2019xts} read\footnote{The first of these two bounds actually originates from two independent constraints, namely $\WC{\Ftwo^2}{}+2\WC{\Ffour}{} - 2\WC{C}{} > 0$ and  $\WC{\Ftwo^2}{}+2\WC{\Ffour}{} + 2\WC{C}{} > 0$, whose combination gives the first bound.}
\begin{align}
    \WC{\Ftwo^2}{}+2\WC{\Ffour}{} - 2|\WC{C}{}| &> 0 \, , \label{eq:POS-Bellazzini1} \\
    \WC{\Ffour}{} &> 0 \, . \label{eq:POS-Bellazzini2}
\end{align}
It is however important to stress that most of the literature, including the above works, explicitly excludes the massless graviton in the derivation of positivity bounds. This is partially related to the technical difficulties originating from subtracting the massless graviton pole and treating the \ac{IR} logarithms. In connection to our work, this has two important implications. First, while there has been some effort in addressing these issues, see \eg{}~\cite{Bellazzini:2019xts,Alberte:2020jsk, Herrero-Valea:2020wxz, Caron-Huot:2021rmr, Caron-Huot:2022ugt}, the situation is not settled yet. In particular, since we allow for massless graviton fluctuations, we will observe a logarithmic running in the \ac{IR}, and we will have to discuss how to treat the resulting logarithms in our \ac{RG} flow when defining the Wilson coefficients. Second, and most importantly, it is not obvious that the strict inequalities above also apply to theories with massless degrees of freedom. The expectation~\cite{Alberte:2020jsk,Herrero-Valea:2020wxz,Henriksson:2022oeu} is that such positivity bounds may be violated once gravity is turned on. The violation would result in a weakening (or even a complete removal~\cite{Herrero-Valea:2022lfd}) of the positivity bounds: sums of dimensionless Wilson coefficients which ought to be positive in unitary, gravity-free theories, could actually be slightly negative~\cite{Alberte:2020jsk,Herrero-Valea:2020wxz,Henriksson:2022oeu}. The amount of allowed violations is not precisely settled but, when present~\cite{Herrero-Valea:2022lfd}, it is generally expected to be an $\mathcal{O}(1)$ quantity in appropriate units.

Specifically, within our system given by the action~\eqref{eq:action-intro}, any potential violation of positivity bounds should naturally be Planck-mass suppressed, as long as the dimensionless Wilson coefficients~\eqref{eq:wilson-coeff-prime} are of $\mathcal O(1)$. To illustrate this argument, let us consider photon-photon scattering. With our effective action~\eqref{eq:action-intro}, the two-to-two scattering amplitude in the low energy and forward scattering limit reads, structurally~\cite{Alberte:2020bdz},
\begin{equation}
    \mathcal A \sim - G_N \frac{s^2}{t} + c \, s^2 + \dots \, ,
\end{equation}
where the first term originates from the massless graviton pole, $c$ is a linear combination of the couplings $G_{\Ftwo{}^2}$ and $G_{\Ffour{}}$, and $s, t$ are the standard Mandelstam variables. Positivity bounds are tied to the positivity of $c$. Using our definitions for the dimensionless Wilson coefficients~\eqref{eq:wilson-coeff-prime}, and assuming $\WC{i}{}\sim -\mathcal{O}(1)$, we have that
\begin{equation}\label{eq:planck-suppression}
    \mathcal A \sim - G_N \frac{s^2}{t} - \mathcal O(1) \, G_N^2 \, s^2 + \dots \sim - \frac{s^2}{t \, M_{Pl}^2} - \mathcal O(1) \frac{s^2}{M_{Pl}^4} + \dots \, .
\end{equation}
This entails that, as long as energies do not reach the Planck scale, $s \ll M_{Pl}^2$, any potential violation is suppressed significantly. As a matter of fact, if a second mass scale $M\ll M_{Pl}$ is present, it is conjectured that only a weaker positivity bound needs to be fulfilled~\cite{Alberte:2020bdz},
\begin{equation}
    c \gtrsim - \frac{\mathcal O(1)}{M^2 M_{Pl}^2} \, ,
\end{equation}
compared to which any potential violation in our system is once again suppressed by a factor of $M^2/M_{Pl}^2$.

\subsection{Entropy positivity bounds and electric weak gravity conjecture}

The \ac{WGC} is among the most important and best-understood criteria within the swampland program. In the context of \ac{ST}, it is strictly related to the No Global Symmetries conjecture, but it can also be motivated by the requirement that no black hole remnants are formed, since they might lead to consistency issues at the \ac{EFT} level~\cite{Arkani-Hamed:2006emk}.
In its simplest form, its electric version states that in any consistent $U(1)$-theory coupled to gravity, there must be at least one electrically charged state, whose dimensionless mass-to-charge ratio is bounded by an order-one number,
\begin{equation}\label{eq:WGC}
    m\leq \sqrt{2} Q\,M_{Pl}\,,
\end{equation}
where $Q=qg$ is the unquantized charge of the state with mass $m$, and $g$ is the $U(1)$-gauge coupling.
Equivalently, there must be an electrically charged state whose charge-to-mass ratio is bounded by that of an extremal charged black hole,
\begin{equation}\label{eq:WCG-lo}
    Q/M\leq Q_{extr}/M_{extr}\,.
\end{equation}
At variance with the condition~\eqref{eq:WGC}, the bound above is not universal, rather it is generally modified by higher derivative terms in the effective action, in the gravitational or in the~$U(1)$ sector~\cite{Kats:2006xp,Cheung:2018cwt,Hamada:2018dde}. Specifically, in a generic photon-graviton \ac{EFT}, the condition~\eqref{eq:WGC} becomes
\begin{equation}\label{eq:WCG-corrected}
    Q/M\leq Q_{extr}/M_{extr}\left(1-\frac{\Delta}{M^2}\right)\,,
\end{equation}
where $\Delta$ entails a non-trivial combination of Wilson coefficients, and the extremality parameter is in the range $\xi=\sqrt{1-Q^2/M^2}\in[0,1]$, or in $[0,1/2]$ for black holes with positive specific heat~\cite{Cheung:2018cwt}. Following the derivations in~\cite{Kats:2006xp,Cheung:2018cwt,Hamada:2018dde}, for a theory of the type~\eqref{eq:action-intro} considered in this paper, the combination reads~\cite{Cheung:2018cwt}
\begin{equation}
    \Delta=(1-\xi)^2d_0+20\xi (8\pi G_N) G_{\eulerterm} -5\xi (1-\xi)(16\pi G_N G_{\eulerterm} +G_{CFF})\,,
\end{equation}
with
\begin{equation}
    d_0 = \frac{G_{\Ftwo^2}}{32\pi G_N} + \frac{G_{\Ffour}}{16\pi G_N} -G_{CFF} = 8\pi G_N (\WC{\Ftwo{}^2}{}+2\WC{\Ffour}{} - 2\WC{C}{})\,,
\end{equation}
and must be non-negative for all $\xi$ in the allowed range for the entropy-positivity-bounds in~\cite{Cheung:2018cwt} to be fulfilled. 
We note that despite the Euler coupling $G_{\eulerterm}$ not entering field equations or scattering amplitudes --- at least not trivially --- it can affect off-shell quantities like the black hole entropy. Thus, it can generally impact the family of positivity bounds attached to the condition
\begin{equation}\label{eq:entr-pos-bound}
    \Delta>0\,.
\end{equation}
Nonetheless, $G_{\eulerterm}$ non-trivially drops out of the linear combination of Wilson coefficients~$d_0$, which is strictly related to the electric \ac{WGC}. Indeed, in the case of a highly charged black hole ($\xi\ll1$), the family of positivity bounds~\eqref{eq:entr-pos-bound} is proportional to the extremality condition of charged black holes~\cite{Cheung:2018cwt} and yields the electric \ac{WGC} for black holes in the presence of higher-derivative corrections~\cite{Kats:2006xp}
\begin{equation}\label{eq:WCG-corrected-expl}
    d_0>0\,.
\end{equation}
This is the condition that we will consider in this manuscript. Similarly to positivity bounds, in the presence of gravitational quantum fluctuations, small violations of the \ac{WGC} are allowed and compatible with unitarity and causality~\cite{Henriksson:2022oeu}.

\section{Computing the AS landscape} \label{sec:landscape} 

In this section, we compute the landscape of \acs{EFT} stemming from asymptotically safe photon-gravity flows. The dynamics is encoded in the action~\eqref{eq:action-intro}. For the computation of the beta functions, we will work with its Euclidean counterpart,
\begin{equation}\label{eq:action}
\begin{aligned}
    \Gamma_k = &\int \text{d}^4x \, \sqrt{g} \, \Big[ -\frac{R}{16\pi G_N} + G_{\eulerterm} \, \eulerterm + \Ftwo{} + G_{\Ftwo^2} \, (\Ftwo{})^2 + G_{\Ffour} \, \Ffour{} + G_{CFF} \, C^{\mu\nu\rho\sigma} F_{\mu\nu} F_{\rho\sigma}\Big] \, ,
\end{aligned}
\end{equation}
where all couplings $G_i(k)\equiv \{ G_N,G_{\eulerterm}, G_{\Ftwo^2}, G_{\Ffour},  G_{CFF}\}$ now depend on the \ac{RG} scale~$k$. As remarked in the introduction, we work in the same spirit as in the \ac{EFT} literature and remove all inessential operators across all scales by an appropriate $k$-dependent field redefinition~\cite{Baldazzi:2021ydj, Baldazzi:2021orb} (see \Cref{sec:setup} for additional details). 

\subsection{Defining the Wilson coefficients}\label{sec:def-wilson}

As anticipated in \Cref{sec:land-def}, identifying the \ac{AS} landscape boils down to computing the set of \ac{IR} endpoints ($k\to0$ limit of the \ac{FRG} flow) of all asymptotically safe trajectories~\cite{Basile:2021krr} (cf. \Cref{fig:AS-land}). Each endpoint corresponds to an \ac{EFT} in the landscape, and is uniquely described by the (generally dimensionful) Wilson coefficients
\begin{equation}
    {W}_{G_i}\equiv \lim_{k\to0} G_i(k)\,.
\end{equation}
The flow has to be computed for the dimensionless counterparts of the couplings $G_i(k)$, \ie{},
\begin{equation}
    g(k) = G_N(k) \, k^2 \, , \quad \! g_{\Ftwo^2}(k) = G_{\Ftwo^2}(k) \, k^4 \, , \quad \! g_{\Ffour}(k) = G_{\Ffour}(k) \, k^4 \, , \quad \! g_{CFF}(k) = G_{CFF}(k) \, k^2 \, .
\end{equation}
The flow for these four dimensionless couplings has been computed with the help of the Mathematica package xAct~\cite{xActwebpage, Brizuela:2008ra, Nutma:2013zea} and a well-tested code~\cite{Knorr:2021lll, Knorr:2022ilz, Baldazzi:2023pep}. Some details on the computation are reported in \Cref{sec:setup}. The complete set of beta functions can be found in the accompanying notebook~\cite{SupplementalMaterial}.

In terms of the above running dimensionless couplings, the dimensionless Wilson coefficients in Eq.~\eqref{eq:wilson-coeff-prime} originate from the following limits 
\begin{equation}\label{eq:WilsonCoeffDef-1}
\begin{aligned}
    \WC{\Ftwo^2}{} &= \lim_{k\to0} \frac{g_{\Ftwo^2}(k)}{(16\pi g(k))^2}\equiv \lim_{k\to0} \, \frac{G_{\Ftwo^2}(k)}{(16\pi G_N(k))^2} \, , \\
    \WC{\Ffour^2}{} &= \lim_{k\to0} \frac{g_{\Ffour^2}(k)}{(16\pi g(k))^2} \equiv \lim_{k\to0} \, \frac{G_{\Ffour}(k)}{(16\pi G_N(k))^2} \, , \\
    \WC{C}{} &=  \lim_{k\to0} \frac{g_{CFF}(k)}{16\pi g(k)}\equiv \lim_{k\to0} \, \frac{G_{CFF}(k)}{(16\pi G_N(k))} \, .
\end{aligned}
\end{equation}
These are the dimensionless ratios of couplings that enter scattering amplitudes in our system, and thus the corresponding positivity bounds. However, some caveats and ambiguities remain. 
Hence, prior to computing the \ac{AS} landscape, we need to discuss some aspects of the definition of the Wilson coefficients in our setup:
\begin{itemize}
    \item \noindent\textbf{Wick rotation}: Our computation of beta functions has been performed with Euclidean signature, as is the standard in the field.\footnote{For recent progress in defining flows directly in Lorentzian signature, see~\cite{Fehre:2021eob}.} We thus have to perform a Wick rotation to relate our couplings to the couplings in the literature on positivity bounds and the \ac{WGC}~\cite{Cheung:2018cwt, Bellazzini:2019xts, CarrilloGonzalez:2023cbf}. It is known that the Wick rotation comes with many issues in gravitational theories~\cite{Baldazzi:2018mtl}. We implement the Wick rotation pragmatically by the rule that spacetime curvature tensors get a minus sign, and field strength tensors a factor of the imaginary unit {\bf{i}}. In practice, this only affects the Einstein-Hilbert term and the kinetic term of the photon in our action. As a result of this procedure, our couplings do not have to be modified and can be used directly in all bounds.
    
    \item \textbf{Definition of Wilson coefficients in the presence of logarithms}: Some of the couplings in our system display a logarithmic running originating from massless graviton fluctuations. Such a logarithmic running introduces an ambiguity in the definition of the Wilson coefficients, since one can always redefine the scale in the logarithm, so that $a_1+\log(k^2/k_1^2)=a_2+\log(k^2/k_2^2)$. This makes it necessary to adopt a prescription to subtract the logarithm and thus to define the Wilson coefficients uniquely (see also~\cite{Basile:2021krr}).

    Grounded on this issue, and in anticipation of the results, it turns out useful to introduce the two combinations
    \begin{equation}
    g_{\pm}(k) = \frac{g_{\Ftwo^2}(k) \pm g_{\Ffour}(k)}{2} \, , 
    \end{equation}
    and the corresponding dimensionless Wilson coefficients
    \begin{equation}\label{eq:WilsonCoeffDef-2}
    \WC{\pm}{} = \lim_{k\to0} \frac{g_{\pm}(k)}{(16\pi g(k))^2}\equiv \lim_{k\to0} \, \frac{G_{\pm}(k)}{(16\pi G_N(k))^2} \, .
    \end{equation}
    The underlying reason is that both $g_{\Ftwo^2}{}$ and $g_{\Ffour^2}{}$ run logarithmically in the \ac{IR}, but with opposite signs. Thus, in combining them, only $g_-$ shows a logarithmic running in the \ac{IR}, whereas $g_+$ does not, allowing us to eliminate one of the two logarithmic ambiguities.

    In practice, we will subtract the logarithmic running to obtain the Wilson coefficient~\WC{-}{}. To parameterize the ambiguity, we will fix our reference scale to be the Planck mass and introduce a parameter, \scaleparameter{}, multiplying $G_N$ inside the logarithm:
    \begin{equation}\label{eq:WilsonCoeffDef-3}
    \begin{aligned}
    \WC{-}{} &\equiv \frac{1}{2}(\WC{\Ftwo^2}{}-\WC{\Ffour^2}{}) - \mathcal{N} \, \ln \left[ \scaleparameter{} G_N k^2 \right] \, . 
    \end{aligned}
    \end{equation}
    In computing the landscape, $\mathcal{N}$ has to be chosen appropriately to remove the universal \ac{IR} logarithmic running. This will be shown explicitly in \Cref{sec:IRparam}, cf.~\eqref{eq:logsubtractioninWC}. Different choices of \scaleparameter{} then correspond to different prescriptions to subtract the logarithm and define the Wilson coefficient~\WC{-}{}.
    
    \item \textbf{Role of Euler coupling}: In four dimensions, the Euler term~\eulerterm{} is topological, and thus the corresponding coupling $G_\eulerterm$ does not enter the right-hand side of the flow equation in conventional treatments. There are two consequences to this. First, the coupling will generically not show a fixed point~\cite{Falls:2020qhj}, but run off to $\pm\infty$ for $k\to\infty$. Modulo the reconstruction problem~\cite{Manrique:2009tj, Morris:2015oca, Fraaije:2022uhg}, this suggests that certain topologies are preferred in the Euclidean path integral, or that only topologies with vanishing Euler term contribute to the Lorentzian path integral~\cite{Knorr:2021slg}. Second, even disregarding the first issue, without any additional constraint we cannot uniquely define the Wilson coefficient of the Euler coupling, as it can be shifted by an arbitrary amount while still fulfilling the same \ac{RG} equation. This problem is exacerbated by the fact that the coupling also runs logarithmically in the \ac{IR}, giving rise to the same ambiguity as for $g_-$. This is generally not a problem, as the Euler coupling typically does not enter scattering amplitudes in four dimensions. Nonetheless, as we shall see, the above issues lead to ambiguities at the level of the bounds involving off-shell quantities.
\end{itemize}

With these caveats in mind, in the next subsections, we will continue our discussion by investigating the analytical properties of two important limits of the \ac{FRG} flow: the fixed point structure in the \ac{UV}, and its universal behavior in the \ac{IR}.

\subsection{Fixed points, UV behavior, and free parameters}

Obtaining the \ac{AS} landscape requires in the first place a \ac{UV} fixed point from which the flow emanates. The fixed points of the \ac{RG} flow are given by the zeros of the vector field of beta functions $\vec \beta(\vec g)$ of all dimensionless couplings $\vec g$. Along with the fixed points, the leading-order behavior of the flow about a fixed point contains information about the predictivity of the corresponding theory, which, in turn, is related to the dimensionality of the corresponding sub-landscape. Indeed, linearizing the beta functions about a fixed point~$\vec g^{\,\ast}$ and solving the flow yields the leading-order scaling
\begin{equation}\label{eq:lin-flow}
    \vec g = \vec g^{\,\ast} + \sum_i \, c_i \, \vec e_i \, \left(\frac{k}{k_0}\right)^{-\theta_i} \, ,
\end{equation}
where the $\vec e_i$ are the unit eigenvectors of the stability matrix $\partial\vec\beta/\partial \vec g$, the $c_i$ are integration constants, $k_0$ is an arbitrary reference scale, and the critical exponents $\theta_i$ are defined as minus the eigenvalues of the stability matrix defined above. The number of relevant directions equates to the number of positive critical exponents and measures the predictivity of the theory generated by the \ac{UV} fixed point: \ac{RG} trajectories reaching the fixed point in the limit~$k\to\infty$ are those whose integration constants $c_i$ multiplying a positive power of $k$ are zero. Thus, the fewer the number of relevant directions for a fixed point, the fewer the number of integration constants to fix to identify an \ac{RG} trajectory, resulting in a higher predictive power.

For asymptotically safe trajectories, the number of non-vanishing integration constants equates to the number $N$ of relevant directions of the flow. We note however that one of the integration constants associated with the relevant couplings can be absorbed to redefine the reference scale,
\begin{equation}
    k_T \equiv c_i^{1/\theta_i} k_0 \, .
\end{equation}
The possibility of redefining the reference scale, or, equivalently, adding an arbitrary shift to the \ac{RG} time $t=\log(k/k_0)$ is related to the invariance of the flow under shifts of the \ac{RG} time $t$. At the same time, $k_T$ has the interpretation of a transition scale from the fixed point (or, \ac{QG}) regime to the \ac{IR} scaling. Indeed, as we shall see in this section, in a gravitational theory $k_T$ is also related to the mass scale defining the Newton coupling. 

In any theory with dimensionful couplings, one needs at least one arbitrary unit mass scale with respect to which other dimensionful quantities can be measured: one of the integration constants of the full theory can be eliminated to define such a scale. Thus, contrary to the common folklore that $N$ relevant directions correspond to $N$ free parameters to be measured to fix the theory, the need to have a unit of measure reduces the number of free parameters to $N-1$ \textit{dimensionless} quantities~\cite{Basile:2021krr}. This number also defines the dimensionality of the landscape in the space of dimensionless Wilson coefficients. The only exception to this argument are scale-free theories. If our universe were a conformal field theory, then all measurable quantities would be dimensionless, in which case the need to introduce one arbitrary mass unit scale would disappear. 

Following this general discussion, in the next subsection we shall present the fixed point structure for our system.

\subsection{Fixed point structure}

The beta functions for the dimensionless couplings $\{g, g_+, g_-, g_{CFF}\}$ are rational functions, and their fixed points correspond to their common zeros. Here we are excluding the Euler coupling $g_\eulerterm$ from the set, since by construction its beta function can only vanish if all other couplings conspire so that $\beta_{g_{\eulerterm}}=0$. This is because, as mentioned in \Cref{sec:def-wilson}, the Gauss-Bonnet operator \eulerterm{} is a topological invariant and thus the Euler coupling cannot enter the right-hand side of the flow equation. As a consequence, $g_{\eulerterm}$ cannot influence the flow of the other couplings, nor of itself, and hence it generically has no fixed points~\cite{Falls:2020qhj, Knorr:2021slg}.

\begin{table}
\begin{center}
\begin{tabular}{|l||l|l|l|l|l|}
\hline
    & $g^*$ & $g_+^*$ & $g_-^*$ & $g_{CFF}^*$ & $\theta_i$                                         \\ \hline \hline
GFP & 0     & 0       & 0       & 0           & $\{-4,-4,-2,-2\}$                                  \\ \hline
MFP & 0     & -12.577 & -10.383 & -0.0901     & $\{4.227, -0.477, -0.723, -1.041\}$                \\ \hline
FP1 & 0.131 & 0.351   & 3.327   & 0.00375     & $\{1.845 , -0.239 \pm 0.0155 \mathbf{i}, -0.291\}$ \\ \hline
FP2 & 0.126 & -0.308  & 4.001   & -0.00410    & $\{1.936, 0.184, -0.141,-0.236\}$                  \\ \hline
\end{tabular}
\end{center}
\caption{Overview of relevant fixed points (those that are both \ac{UV}-complete and connected to the \ac{GFP} in the \ac{IR}), with their coordinates $g_{i}^\ast$ in theory space and their critical exponents $\theta_i$.}\label{tab:FP}
\end{table}

The fixed points of our system that are reliable in our approximation are reported in \Cref{tab:FP}, together with the corresponding critical exponents. Despite the beta functions being rational functions, the search for fixed points generally requires resorting to numerical methods. Yet, when turning off Newton's coupling, $g=0$, one can find the zeros of the remaining three beta functions analytically. This identifies the ``pure'' \acp{MFP}. Next to the standard \ac{GFP} (first entry in \Cref{tab:FP}), with critical exponents~$\theta$ given by the classical mass dimensions of the couplings, we find two real \acp{MFP}. The first one is denoted by ``MFP'' in \Cref{tab:FP}, and has a single relevant direction. The second one has highly non-canonical critical exponents, making it unreliable in our approximation, and it is also shielded from the \ac{GFP} by a singularity in the beta functions, meaning that it cannot have a standard \ac{IR} behavior. We thus excluded this second matter fixed point from~\Cref{tab:FP}.

To search for fully interacting fixed points, we employed a numerical fixed point search together with a random grid of starting points for the Newton-Raphson iteration, focusing on the region close to the \ac{GFP}. For this, we first used a hypercube around the \ac{GFP} with~$10^6$ randomly distributed starting points. We then successively enlarged the cuboid in all directions, except the $g$-direction, which we restricted to the interval $[0,2\pi]$. This is because, on the one hand, we are only interested in fixed points with positive Newton's coupling, and on the other hand, there is a singular line at $g=2\pi$ in the flow, beyond which fixed points cannot be connected to a standard \ac{IR} behavior.\footnote{This should be understood as a maximum value for the first singularity encountered when increasing $g$ from 0 to positive values, and it stems from our renormalization condition for $\lambda$, see \Cref{sec:setup}.} The largest cuboid that we investigated was $g\in[0,2\pi]$, $g_\pm\in[-100\pi^2,100\pi^2]$, $g_{CFF}\in[-100,100]$.\footnote{The factors of $\pi$ are chosen for convenience --- rescaling the couplings $\{g,g_\pm,g_{CFF}\}\to\{\pi \tilde g,\pi^2 \tilde g_\pm, g_{CFF}\}$ removes all factors of $\pi$ in the beta functions.}

With this method, we identified two reliable fixed points that are connected to the \ac{GFP} in the \ac{IR}. Both are displayed in \Cref{tab:FP}. The first one (FP1) comes with a single relevant direction, and thus yields the most predictive theory. By contrast, the second fixed point (FP2) has two relevant directions, with a peculiar magnitude of the two relevant critical exponents: $\theta_1$ is an order of magnitude larger than $\theta_2$. We anticipate here that this makes the computation of the sub-landscape resulting from this fixed point numerically challenging. 
In agreement with the discussion above and previous computations in \ac{AS}~\cite{Knorr:2021slg, Knorr:2022ilz, Knorr:2023usb, Baldazzi:2023pep}, we find that
\begin{equation}
    k \partial_k g_\eulerterm(k) > 0 \, ,
\end{equation}
at both FP1 and FP2. This entails that the coupling $g_\eulerterm$ runs to $+\infty$ for $k\to\infty$, and does not have a common fixed point with the other couplings, as expected.

To summarize, our system has three fixed points that can serve as a \ac{UV} completion of the theory, and that are connected to the \ac{GFP} ($g_i^\ast=0$) in the \ac{IR}.\footnote{For general theories including couplings with positive mass dimension (\eg{}, a cosmological constant), the condition of a \ac{UV} fixed point being connected to the \ac{GFP} generalizes to the existence of \ac{RG} trajectories departing from it and reaching or passing close to the \ac{GFP} in the \ac{IR}.} For the case of fully interacting fixed points (FP1 and FP2), this is needed to recover \ac{GR} at low energies. The \ac{MFP}, lying at $g^\ast=0$, instead corresponds to a gravity-free \ac{UV} completion of self-interacting photons. In the next subsections, we will show how to parameterize the \ac{IR} behavior of the flow, which is key to compute the sub-landscapes stemming from the three possible \ac{UV} completions presented in this subsection. In particular, an important role will be played by the so-called ``separatrices'' --- critical \ac{RG} trajectories connecting couples of fixed points and separating different behaviors of the flow. In the next subsections, we will describe the sub-landscapes identified by the three interacting fixed points in \Cref{tab:FP}, as well as the global geometry of the \ac{AS} landscape.

\subsection{Computing the landscape: parameterizing the IR behavior of the flow}\label{sec:IRparam}

Parameterizing the \ac{IR} behavior of the beta functions is a necessary step to efficiently compute the \ac{AS} landscape. The strategy is to expand the beta functions in the \ac{IR} about the \ac{GFP} by using the ans\"atze $a (k/k_0)^{-d_c}(b+c \log(k/k_{ref}))$ for the running of the couplings, with $d_c$ being the classical mass dimension of the corresponding coupling. Inserting these into the beta functions, one then extracts the coefficients $a,b,c$. This procedure yields the following scalings in the limit $k\to0$, which are \ac{IR}-universal, \ie{}, independent of the specific \ac{UV} completion:
\begin{subequations}
    \begin{align}
        g(k) &\simeq g_{IR} \left(\frac{k}{k_0}\right)^2\,,\\
        g_+(k) &\simeq g_{IR}^2\left(\frac{k}{k_0}\right)^4 f_+ \, , \\
        g_-(k) &\simeq g_{IR}^2\left(\frac{k}{k_0}\right)^4\left(f_- - \frac{548}{15}\ln\left[c_l \, g_{IR}\left(\frac{k}{k_0}\right)^2\right]\right) \, , \label{eq:logsubtractioninWC} \\
        g_{CFF}(k) &\simeq g_{IR}\left(\frac{k}{k_0}\right)^2 f_c \, .
    \end{align}
\end{subequations}
In this, $c_l$ parameterizes the ambiguity in defining the argument of the logarithm, in accordance with the discussion in \Cref{sec:def-wilson}. The remaining parameters $\{f_\pm, f_c\}$ are to be determined by the \ac{RG} flow: they depend on the fixed point one starts with and on the specific \ac{RG} trajectory departing from it. As the running dimensionful Newton coupling is $G(k)\equiv g(k)k^{-2}$, the combination $g_{IR} k_0^{-2}$ \textit{defines} the Newton coupling as
\begin{equation}
    G_N\equiv \lim_{k\to0}g(k)k^{-2}=g_{IR} k_0^{-2}\,.
\end{equation}
In turn, $G_N$ sets the scale of \ac{QG}, and thereby provides a scale with respect to which all other dimensionful quantities can be meaningfully defined. Indeed, while the unit scale $G_N$ is arbitrary, all other Wilson coefficients can be defined by the dimensionless ratios~\eqref{eq:WilsonCoeffDef-1} and~\eqref{eq:WilsonCoeffDef-2} --- the only quantities that are relevant to assess positivity bounds.

\subsection{Sub-landscape from FP1}

The most predictive \ac{UV} completion in our setup is provided by FP1, as it comes with one relevant direction only. The free parameter attached with its only relevant direction fixes the scale of \ac{QG}, $G_N$, and thereby also fixes the units to define all other Wilson coefficients. All remaining dimensionless Wilson coefficients are thus predicted uniquely.
\begin{figure}
    \centering
    \includegraphics[width=0.55\textwidth]{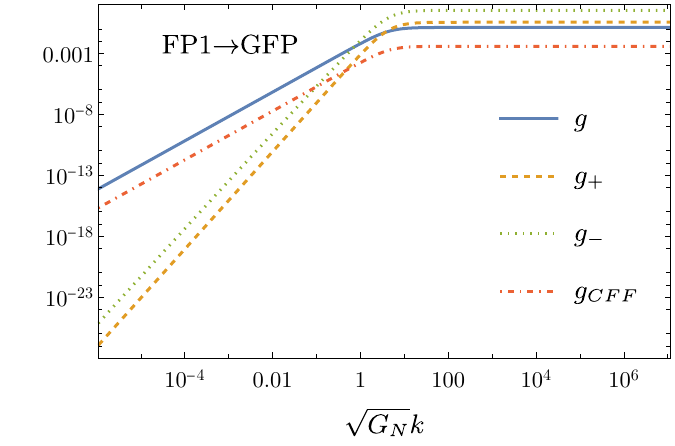}
    \caption{Flow of all dimensionless couplings along the only \ac{RG} trajectory emanating from FP1 and reproducing \ac{GR} at low energies, \ie{}, the separatrix joining FP1 and \ac{GFP}. The running is shown as a function of the \ac{FRG} scale $k$ in Planck units, with a logarithmic scale on both axes. One can clearly see the scaling regimes of the two fixed points: for $k \gtrsim G_N^{-1/2}$, the couplings flow towards their fixed point values, whereas for $k \lesssim G_N^{-1/2}$, they run according to their mass dimensions.}
    \label{fig:separatrix}
\end{figure}

The single relevant direction and the presence of an \ac{IR} fixed point (the \ac{GFP}) imply that there are only two trajectories emanating from FP1: one diverging, and one approaching the \ac{GFP}, as $k\to0$. The latter trajectory (or, more precisely, the family of trajectories for different values of the \ac{QG} scale) corresponds to the separatrix between FP1 and the \ac{GFP} in theory space. The flow of the dimensionless couplings along this separatrix is shown in \Cref{fig:separatrix}.

The sub-landscape stemming from FP1 is thus a single point, and it corresponds to the $k\to0$ limit of the FP1-\ac{GFP} separatrix. The separatrix can be computed numerically by integrating the beta functions with initial conditions close to FP1, perturbed in the direction of the eigenvector corresponding to the positive critical exponent. The only \ac{EFT} in the sub-landscape is identified by the following Wilson coefficients:
\begin{equation}\label{eq:landscapeFP1_1}
    \WC{+}{} = 0.00792 \, , \quad 
    \WC{C}{} = 0.000550 \, . 
\end{equation}
The Wilson coefficient of~\WC{-}{} suffers the aforementioned ambiguity due to its logarithmic running. Trivially, this coefficient depends logarithmically on the parameter~\scaleparameter{}. For~$\scaleparameter{}=16\pi$, we obtain
\begin{equation}\label{eq:landscapeFP1_2}
    \WC{-}{} = 0.0955 \, . 
\end{equation}
We shall use these Wilson coefficients in the next section to investigate the compatibility with positivity bounds and the \ac{WGC}.

\subsection{Sub-landscape from FP2}

We move on to discuss the sub-landscape connected to FP2. This fixed point has two relevant directions. Following from the earlier discussion that one of the free parameters sets the \ac{QG} scale, the ensuing sub-landscape is one-dimensional: it is a line in the space of dimensionless Wilson coefficients. Here, all \ac{RG} trajectories fall into three classes: those that approach the \ac{GFP} in the \ac{IR}, those that diverge, and the two boundary trajectories between these two behaviors. The first set makes up the sub-landscape.

Before we discuss our results, let us briefly point out a technical difficulty. Since the positive critical exponents of this fixed point are one order of magnitude apart, see \Cref{tab:FP}, the numerical determination of the sub-landscape requires high precision.\footnote{Specifically, a working precision of 32 digits was required for our data points when using Mathematica's NDSolve routine. We then chose initial conditions on a small circle about FP2 within its critical hypersurface, parameterized by an angle. We picked one specific trajectory that is well within the sub-landscape. From there, we changed the angle by a step of $\pi/10$. If the successive trajectory is still flowing into the \ac{GFP}, we continue the procedure by changing the angle by the same amount, otherwise, we discard the trajectory, half the step size, and compute the next trajectory. We repeated this procedure until the step size fell below $10^{-18}\pi$. Finally, we added some additional trajectories to improve the density of data points in the sub-landscape.} Trajectories that are close to each other near the fixed point can vastly differ in their \ac{IR} physics. In practice, this issue is realized in different ways within the sub-landscape attached to FP2. This is because the above-mentioned boundary trajectories are separatrices that connect FP2 to FP1 and \ac{MFP}, respectively.\footnote{This means that the part of theory space that is both \ac{UV}-complete and connected to the \ac{GFP} in the \ac{IR} looks very similar to that of the system of a shift-symmetric scalar field coupled to gravity in the same order of the derivative expansion, see~\cite{Knorr:2022ilz}.} As a consequence, going towards the separatrix to FP1, the Wilson coefficients approach those of FP1. On the other hand, approaching \ac{MFP}, the Wilson coefficients become larger and larger. The reason is that one approaches the separatrix between \ac{MFP} and the \ac{GFP}. For this trajectory, Newton's coupling vanishes identically, so our definition of Wilson coefficients is ill-defined and only their ratios are still well-defined, cf. \Cref{sec:MFP-land}.

\begin{figure}
    \centering
    \includegraphics[scale=0.7]{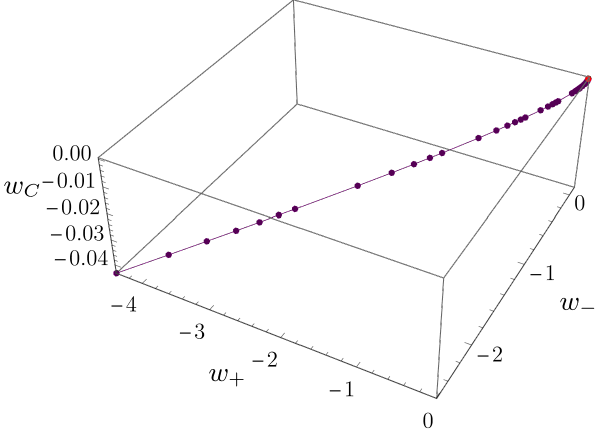}\hspace{0.6cm}\includegraphics[scale=0.65]{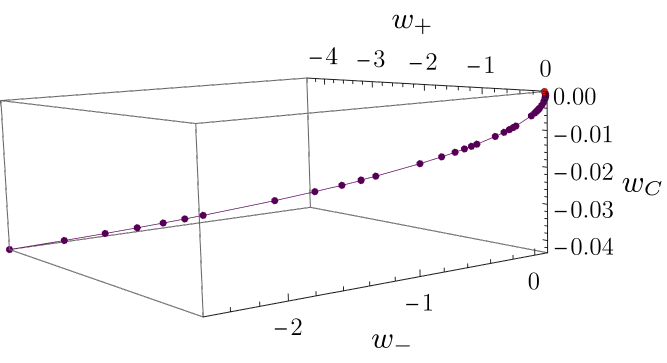}\\\includegraphics[scale=0.7]{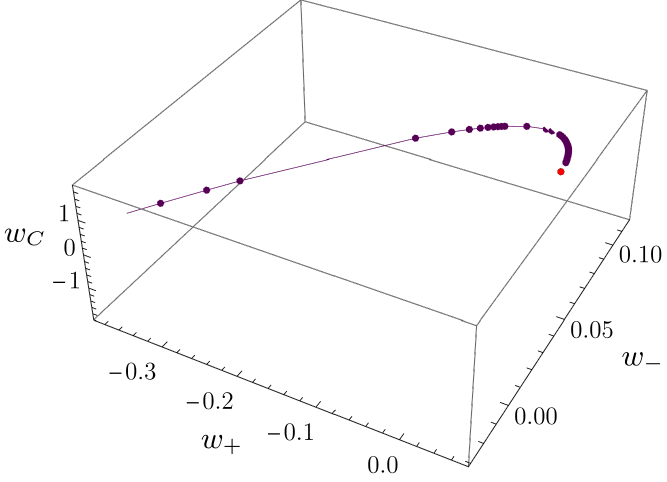}
    \caption{Different views and zooms on the \ac{AS} landscape, in the space of dimensionless Wilson coefficients~$\{\WC{+}{},\WC{-}{},\WC{C}{}\}$. The dots denote the Wilson coefficients computed as the \ac{IR} limit of a sample of asymptotically safe \ac{RG} trajectories. The purple dots are the \ac{IR} coefficients arising from FP2, whilst the red one is the sub-landscape stemming from FP1. The line connecting the points is an interpolating function. The overall landscape, given by the union of the sub-landscapes from FP1 and FP2, is approximately a straight line. The line however bends in the proximity of the FP2-landscape, forming a tiny ``candy cane'' which continuously connects the two sub-landscapes. Moreover, the entire landscape falls approximately onto a plane. Both the near-flatness, already encountered in~\cite{Basile:2021krr}, and the contiguity of the sub-landscapes are non-trivial features.}
    \label{fig:3DlandscapeFP2}
\end{figure}

With this remark out of the way, let us discuss the sub-landscape of FP2. In \Cref{fig:3DlandscapeFP2} we show different views on the Wilson coefficients within this sub-landscape (purple dots), as well as the \ac{EFT} generated by FP1 (red dot). The overall geometry resembles that of a stretched-out candy cane: most of this sub-landscape is approximated by a straight line (the extended ``strabe''). At one end, this extended strabe terminates at the sub-landscape of \ac{MFP}, which we will discuss in the next subsection. At the other end, there is a small curved part (the ``warble'') that connects it to the sub-landscape from FP1, and gives the characteristic candy cane shape.

A thorough analysis of our $96$ data points shows that the extended strabe part of the sub-landscape can be approximated by a square root plus linear fit in the $\{\WC{+}{},\WC{-}{}\}$-plane, and by a quadratic fit along the ~$\WC{C}{}$-direction.\footnote{The form of these functions was selected by the fact that they fit large parts of the sub-landscape, significantly beyond the selection of data points used to obtain them.} Concretely, fitting the $10$ data points that lie furthest away from the warble region, we find
\begin{equation}\label{eq:candy-cane-pm-linear}
    \WC{-}{} \approx 0.8022 \WC{+}{} + 0.4353 \sqrt{-\WC{+}{}} + 0.03453 \, ,
\end{equation}
and
\begin{equation}\label{eq:candy-cane-pmC-quadratic}
\begin{aligned}
        \WC{+}{} &\approx - 1666 \WC{C}{2} + 32.78 \WC{C}{} + 0.1566 \, , \\
        \WC{-}{} &\approx - 1359 \WC{C}{2} + 5.930 \WC{C}{} + 0.2341 \, .
\end{aligned}
\end{equation}
These fits are valid in the limit of large Wilson coefficients, \ie{} as $\WC{\pm,C}{}\to-\infty$, so that, to leading order, the extended strabe is a straight line in $\{\WC{\pm}{}, \WC{C}{2}\}$. The rather large coefficients in the quadratic fits are entirely due to the fact that $\WC{C}{}$ is much smaller than $\WC{\pm}{}$ for our data points.

\begin{figure}
    \centering
    \includegraphics[scale=0.7]{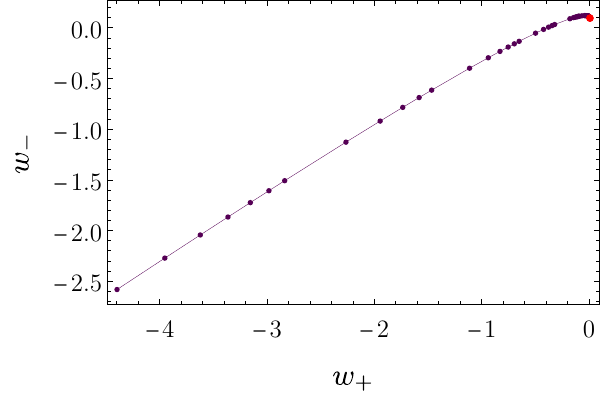}\hfill\includegraphics[scale=0.7]{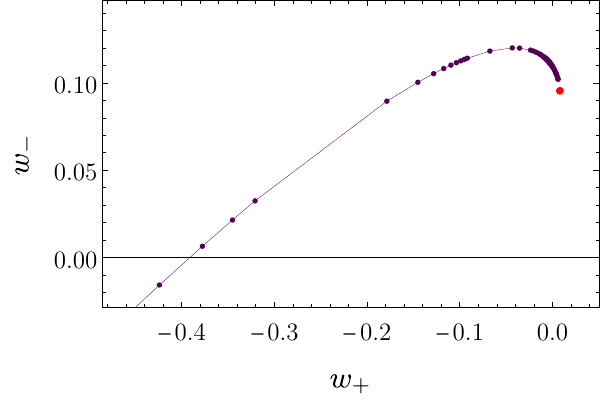}
    \caption{Projection of the \ac{AS} landscape onto the bidimensional theory space~$\{\WC{+}{},\WC{-}{}\}$. In this plane the landscape preserves the characteristic ``candy cane'' shape. This is because the third Wilson coefficient varies very weakly across the entire landscape, $\WC{C}{}\in[-0.043,0.00026]$. As in \Cref{fig:3DlandscapeFP2}, purple dots are \acp{EFT} coming from FP2, whereas the red dot is the only \ac{EFT} associated with FP1.}
    \label{fig:2DlandscapeFP2}
\end{figure}

Remarkably, the \emph{entire} sub-landscape originating from FP2 approximately (and thus the whole \ac{AS} landscape) lies on a plane described by
\begin{equation}\label{eq:FP2_landscape_plane}
    \WC{C}{} = 0.0049155 + 0.038755 \WC{+}{} - 0.047318 \, \WC{-}{} \, .
\end{equation}
For the points of the sub-landscape that were computed, the distance from this plane does not exceed $0.000057$. Moreover, as~$\WC{C}{}$ covers the small range $[-0.043,0.00026]$ for our data points, the shape of the sub-landscape is approximately preserved when projecting it onto the bidimensional space~$\{\WC{+}{},\WC{-}{}\}$, see \Cref{fig:2DlandscapeFP2}. We remark once again that to determine all the above relations, we used $\scaleparameter{}=16\pi$ to fix the logarithmic ambiguity in~$\WC{-}{}$.

Similar findings were also obtained in the first paper that computed a landscape from the \ac{RG} flow of an asymptotically safe gravitational theory~\cite{Basile:2021krr}. Concretely, in~\cite{Basile:2021krr} the one-loop flow of quadratic gravity was investigated, and the resulting two-dimensional landscape was approximately a plane. This is an intriguing and highly non-trivial feature that deserves further investigation.

\subsection{Sub-landscape from MFP}\label{sec:MFP-land}

The fixed point \ac{MFP} lies at $g=0$ and thus it involves no gravity. Nonetheless, since it acts as one of the boundaries of the sub-landscape attached to FP2, we will briefly discuss the properties of its sub-landscape. First of all, since $g=0$ at the fixed point, the separatrix between \ac{MFP} and the \ac{GFP} will have $g=0$ at all scales, corresponding to a pure matter theory. This also entails that we have to use a different dimensionful coupling to set the scale. Second, and intriguingly, on this separatrix, we have an \emph{exact} relation between $g_+$ and $g_-$,
\begin{equation}\label{eq:MFP_sep_pmrelation}
    \frac{g_-}{g_+} = y \approx 0.826 \, , 
\end{equation}
where $y$ is the real root of the polynomial $(-73+131y-59y^2+9y^3)$. This relation is fulfilled along the whole trajectory, which reduces the number of independent couplings to two. Stated differently, the Wilson coefficient resulting from the ratio $g_-/g_+$ is exactly $y$. Finally, since the fixed point has one relevant direction, the sub-landscape has only one non-trivial dimensionless Wilson coefficient. We find that
\begin{equation}\label{eq:MFP_WC}
    \frac{G_{CFF}}{\sqrt{-G_+}} \approx -0.0290 \, .
\end{equation}
We were able to compute its exact value in a closed form as well. Since only limited insights can be gained from it, we present it in~\eqref{eq:MFP_exact_WC} in \Cref{sec:MFP_WC}.

The importance of the relations~\eqref{eq:MFP_sep_pmrelation} and~\eqref{eq:MFP_WC} derives from the fact that, since the sub-landscape of FP2 is bounded by \ac{MFP}, the relations must be fulfilled at the asymptotic end of the extended strabe. For our data points, this is not yet fulfilled --- for the last data point, we find
\begin{equation}
    \frac{g_-}{g_+} \approx 0.5997 \, , \quad \frac{g_{CFF}}{\sqrt{-g_+}} \approx -0.02068 \, .
\end{equation}
This simply signals that our data at the open end of the strabe are not yet in the asymptotic regime, so they do not yet match the scaling of \ac{MFP}. This emphasizes the aforementioned numerical challenge of mapping out the full sub-landscape of FP2.

We can nevertheless try to extract the exact limits~\eqref{eq:MFP_sep_pmrelation} and~\eqref{eq:MFP_WC} from our fits of the extended strabe, Eq. \eqref{eq:candy-cane-pm-linear} and~\eqref{eq:candy-cane-pmC-quadratic}. While using the square root plus linear fit~\eqref{eq:candy-cane-pm-linear} yields~$y\approx 0.8022$, employing the ratio of quadratic fits~\eqref{eq:candy-cane-pmC-quadratic} and sending~$\WC{C}{}\to\infty$, we get~$y\approx 0.816$. Both values are very close to the exact value. Likewise, using the quadratic fit for~$\WC{+}{}$ to compute the ratio~\eqref{eq:MFP_WC}, we obtain $-0.0245$, which differs from the exact value by about $16\%$. Last but not least, we can also use the plane equation~\eqref{eq:FP2_landscape_plane} to estimate~\eqref{eq:MFP_sep_pmrelation}, which is given by the ratio of the coefficients in front of~$\WC{\pm}{}$. From this, we get $y\approx0.819$. This lends more support to the idea that the entire sub-landscape of FP2, including the asymptotic region close to \ac{MFP}, indeed approximately lies on a plane.

\section{Comparing the AS landscape with positivity, entropy, and weak gravity bounds}\label{sec:comparisonIR}

\begin{figure}
    \centering
    \includegraphics[scale=0.7]{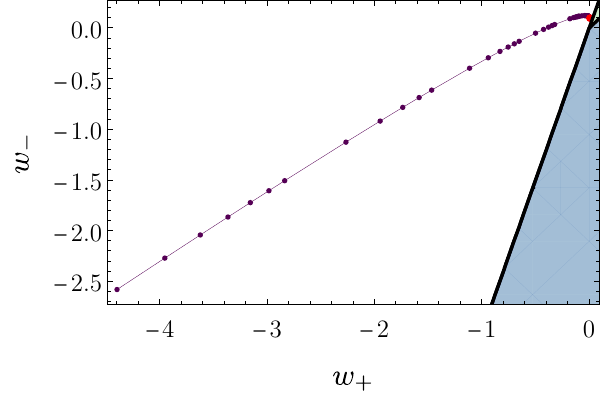}\hfill\includegraphics[scale=0.7]{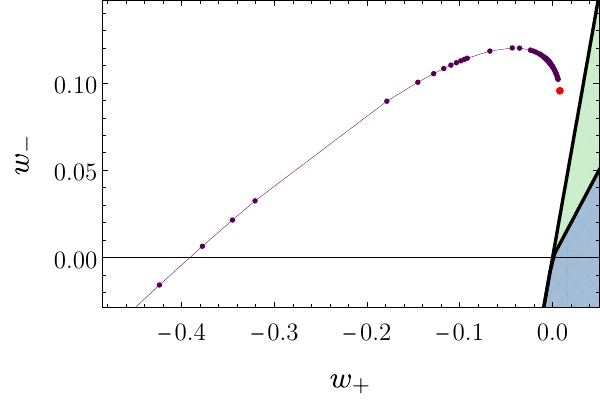}
    \caption{Comparison of the \ac{AS} landscape with the combined positivity bounds (blue region) and \ac{WGC} (green region). The two bounds partially overlap in this projection, with the first combined bounds being stronger than the second. The \ac{AS} landscape lies outside of the region where the bounds are satisfied, and the violation is minimized by the sub-landscape identified by FP1 (red dot).}
    \label{fig:2DlandscapeFP2_vs_posbounds}
\end{figure}

\begin{figure}
    \centering
    \includegraphics[scale=0.7]{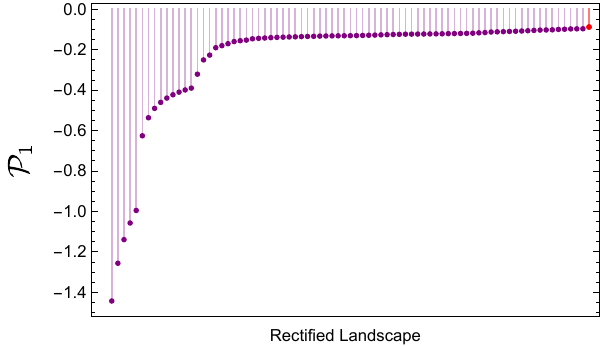}\hfill\includegraphics[scale=0.7]{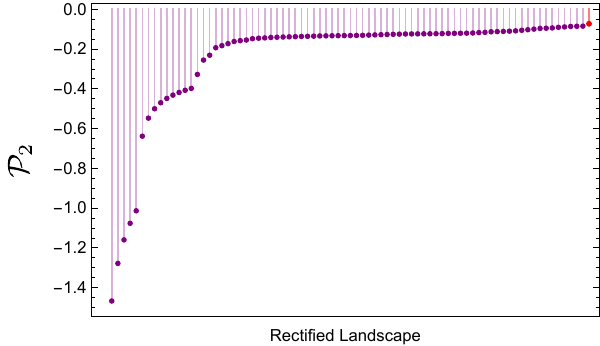}\\ \includegraphics[scale=0.7]{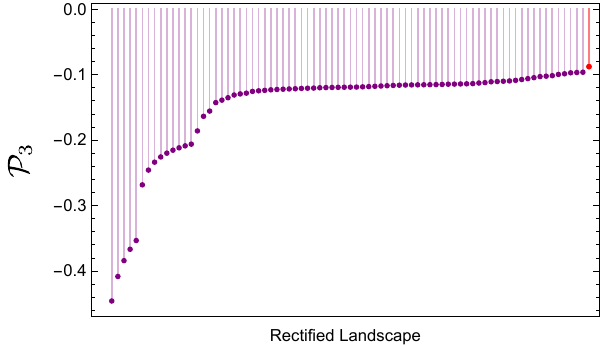}\hfill\includegraphics[scale=0.7]{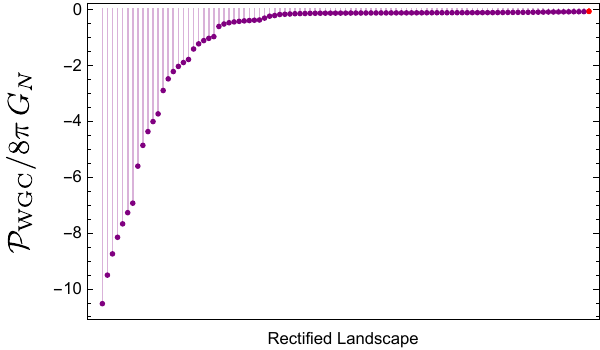}
    \caption{Planck-scale suppressed violations of positivity bounds and \ac{WGC} on the rectified landscape, where data points are equidistant for better representation. As in other figures, the red dot corresponds to the single-point landscape generated by FP1, while the purple dots correspond to the \acp{EFT} in the sub-landscape identified by FP2. The plots show the dimensionless version of the violation for each $\mathcal{P}_i$, \ie{}, their deviation from positivity: the more negative, the larger the violation. As is apparent, the violation is minimized by the \ac{EFT} associated with the most predictive fixed point (red dot) and gets large as one moves away from it along the sub-landscape of FP2.}
    \label{fig:violations-P-bounds}
\end{figure}

\begin{figure}
    \centering
    \includegraphics[scale=0.7]{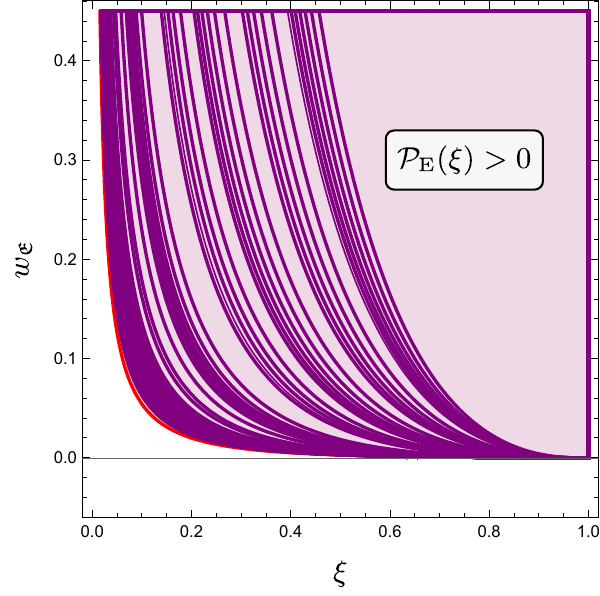}
    \caption{Constraints on the Euler Wilson coefficient resulting from the family of entropy bounds~$\mathcal{P}_E(\xi)$. The bound depends on the point in the landscape: each of the purple lines corresponds to a point in the sub-landscape of FP2, whereas the red line is attached to the sub-landscape of FP1. In order to satisfy the bound, $w_{\eulerterm{}}$ needs to be more positive for \acp{EFT} lying away from the single-point sub-landscape. Moreover, the bound is a function of the extremality parameter~$\xi$. The smaller~$\xi$ is, the stronger the bound. For~$\xi=1$ the bound would imply the weak constraint $\WC{\eulerterm}{}\gtrsim 0$. 
    Requiring that every bound in the family $\mathcal{P}_E(\xi)$ has to hold, implies that the strongest of them has to be satisfied. As is apparent from the figure, the strongest corresponds to the boundary case $\xi=0$: on the one hand, for $\xi=0$ the Euler Wilson coefficient drops out of $\mathcal{P}_E(\xi)$, and the resulting bound is the \ac{WGC}. On the other hand, taking the limit $\xi\to0^+$, independent of the specific \ac{RG} trajectory the positivity of $\mathcal{P}_E(\xi)$ would imply $w_C\to+\infty$. The apparent mismatch is due to the weak violation of the \ac{WGC} across the whole landscape.}
    \label{fig:entropy-bound}
\end{figure}

In terms of the dimensionless Wilson coefficients~\eqref{eq:WilsonCoeffDef-1} and~\eqref{eq:WilsonCoeffDef-2}, the bounds introduced in \Cref{sec:basics} read
\begin{equation}
\begin{aligned}
    & \text{Positivity bounds:}\qquad && \mathcal{P}_1=(16\pi G_N)^2\Lambda^4(2\WC{+}{}-\WC{-}{}-|\WC{+}{}|)>0\,, \\
    & && \mathcal{P}_2=3\WC{+}{}-\WC{-}{}-2|\WC{C}{}|>0\,,\\
    & && \mathcal{P}_3=\WC{+}{}-\WC{-}{}>0\,,\\
    & \text{Entropy bound:} && \mathcal{P}_E= 8\pi G_N ( (1-\xi)^2 (3\WC{+}{}-\WC{-}{})\\
    & && \phantom{\mathcal{P}_E}-2(1-\xi)(1+4\xi)\WC{C}{}+10\xi(1+\xi)\WC{\eulerterm}{})>0\,,\\
    & \text{WGC:} && \mathcal{P}_{\text{WGC}}=8\pi G_N (3\WC{+}{}-\WC{-}{}-2\WC{C}{})>0\,.
\end{aligned}
\end{equation}
Insomuch as the only scale of our system is the Planck mass, we will set~$\Lambda^2=1/(16\pi G_N)$, so that in the first bound all multiplicative factors disappear. At this point, it is important to highlight that while the individual Wilson coefficients $\WC{i}{}$ are neither gauge/parameterization independent, nor invariant under field redefinitions, the combinations of them appearing in scattering amplitudes and also in the $\mathcal{P}_i$ are.
In this section we will compare the above conditions with the \ac{AS} landscape derived in the previous section. Our results are summarized in \Cref{fig:2DlandscapeFP2_vs_posbounds}, \Cref{fig:violations-P-bounds}, and \Cref{fig:entropy-bound}.

The comparison of the global \ac{AS} landscape of our system with the positivity bounds~$\mathcal{P}_i$ and the \ac{WGC} condition~$\mathcal{P}_{\text{WGC}}$ shows that the arguments and conjectures in~\cite{Alberte:2020jsk,Herrero-Valea:2020wxz,Henriksson:2022oeu} are strictly realized within our system: Planck-scale suppressed violations of both positivity bounds and \ac{WGC} occur across the entire landscape.\footnote{Whether positivity bounds are strictly fulfilled in some parts of the landscape may also depend on the approximations employed and, in truncated systems, on the choice of gauge and parameterization to set up the flow~\cite{positivityastrid}.}
This is visually evident in \Cref{fig:2DlandscapeFP2_vs_posbounds}, where we show the projection of the \ac{AS} landscape onto the plane $\{\WC{+}{},\WC{-}{}\}$, together with the regions where positivity bounds and the \ac{WGC} hold. We need to emphasize though that the theory space is parameterized via dimensionless Wilson coefficients so that all numbers are expressed in units of the Newton coupling, cf. Eq.~\eqref{eq:WilsonCoeffDef-1} and~\eqref{eq:WilsonCoeffDef-2}. 

Focusing on the violation of the \ac{WGC}, the comparison of our results with that in~\cite{Basile:2021krr} indicates the importance of the $U(1)$ sector and non-perturbativity in assessing the validity of the \ac{WGC}. Indeed, \cite{Basile:2021krr} assumed electromagnetic duality and investigated the intersections between the \ac{AS} landscape in one-loop quadratic gravity~\cite{Codello:2006in} and some swampland conjectures, including the \ac{WGC}. The \ac{AS} landscape of~\cite{Basile:2021krr} was fully located inside the region $\mathcal P_{WGC}>0$, so that the \ac{WGC} was strictly valid throughout the landscape. The deviation of our result from that in~\cite{Basile:2021krr} is to be attributed to several factors, from the improved computational setup (full-fledged \ac{FRG} versus perturbative computation), to the inclusion of the $U(1)$ sector with essential couplings only, and the different universality class of the \ac{UV} fixed point~\cite{Codello:2006in}.

The dimensionless amount of violation for each $\mathcal{P}_i$ is shown in \Cref{fig:violations-P-bounds}. Each dot quantifies the dimensionless deviation from positivity of $\mathcal{P}_i$ at a point of the rectified version of the landscape, where, aiming at a better visualization, all data points have been deformed to be equidistant. The more negative $\mathcal{P}_i$ is, the larger the violation. In particular, the only \ac{EFT} predicted by FP1 --- denoted by a red dot --- is the one where the violation is smallest. Moving along the landscape, and away from the red dot, the violation gets larger and larger. 

Let us also briefly discuss the positivity bounds for the sub-landscape deriving from \ac{MFP}. Due to the exact expressions for the Wilson coefficients, Eq.~\eqref{eq:MFP_sep_pmrelation} and~\eqref{eq:MFP_WC}, and the fact that $g_+<0$ along the separatrix, we straightforwardly find that all positivity bounds are violated. Since this theory excludes gravitational fluctuations so that standard positivity bounds should apply, this entails that the \ac{UV} completion of self-interacting photons provided by \ac{MFP} is likely not unitary. From this perspective, \ac{AS} might act as a unitarizer of the photonic theory through the fixed points FP1 and FP2.

So far we discussed the comparison of the \ac{AS} landscape with the bounds~$\mathcal{P}_{1,2,3}$ and $\mathcal{P}_{\text{WGC}}$. We now turn our attention to the family of entropy-based positivity bounds~$\mathcal{P}_{\text{E}}(\xi)$. This constraint deserves a separate discussion because of its dependence on an external parameter, and, more notably, due to the appearance of the Euler Wilson coefficient~$w_{\eulerterm{}}$. As already discussed, the Euler coupling is special in purely field-theoretic setups, because its flow typically does not have non-trivial fixed points, and hence the corresponding Wilson coefficient~$w_{\eulerterm{}}$ is unconstrained. This atypical behavior and the lack of \ac{RG}-induced constraints are generally not considered as issues, inasmuch~$w_{\eulerterm{}}$ does not enter on-shell quantities like scattering amplitudes. Yet, the Gauss-Bonnet term impacts off-shell quantities like the entropy~\cite{Myers:1988ze,Myers:1998gt,Clunan:2004tb,Azeyanagi:2009wf,Cheung:2018cwt,Platania:2023uda}, even in four spacetime dimensions. It can consequently enter positivity bounds such as~$\mathcal{P}_{\text{E}}(\xi)$. Such bounds can then result in constraints for the Euler Wilson coefficient, as shown in \Cref{fig:entropy-bound}.
The individual constraints depend both on the extremality parameter~$\xi$ and on the point of the landscape. The bounds become stronger for points of the landscape further away from the single-point landscape and for smaller values of $\xi$. Thus, requiring that they all hold results in the requirement that the strongest of them is satisfied. In our case the strongest of the entropy bounds is realized in the limit $\xi\to0^+$, in which case the Euler Wilson coefficient is constrained to be $w_C\to+\infty$. At the same time, setting $\xi=0$ in the entropy positivity bounds $\mathcal{P}_E(\xi)$ yields the \ac{WGC} and decouples the Euler Wilson coefficient. This discontinuity is due to the slight violation of the \ac{WGC}, since if $\mathcal{P}_{\text{WGC}}$ were positive, the limit $\xi\to0^+$ would have implied $w_C>-\infty$ instead, \ie{}, the absence of a bound.

Once again, this result is to be contrasted with the conclusions drawn in~\cite{Basile:2021krr}. The use of a one-loop approximation in pure quadratic gravity has the effect of generating a fixed point for the Euler coupling, which is thus constrained to be positive in the landscape. In~\cite{Basile:2021krr} this entailed the possibility of testing the validity of the entropy bounds~$\mathcal{P}_E(\xi)$ within the \ac{AS} landscape, in the Stelle universality class~\cite{Codello:2006in}.

Overall, our results point to an intriguing lesson on the role of the Euler coupling in \ac{AS}. Wilson coefficients associated with boundaries or spacetime topology, like the Euler coupling in our case, remain unbounded by the \ac{RG} flow. Conditions like the entropy bounds~$\mathcal{P}_E(\xi)>0$ thus turn into constraints for these couplings. Predictivity then hinges on a better understanding of the role of boundaries in \ac{AS}, and perhaps the necessity of relating bulk and boundary Wilson coefficients via an \ac{AS} version of the holographic principle.

\section{Positivity bounds and WGC beyond Wilson coefficients: flowing conditions}\label{sec:comparisonFlow}

Motivated by the cases of Ward identities~\cite{Litim:1998qi} and the $C$-theorem~\cite{Cardy:1988cwa}, in this section we take a more general perspective, and ask the question whether positivity bounds and the \ac{WGC} should hold at any \ac{FRG} scales $k\geq0$, or if they only apply to fully renormalized quantities. Along these lines, there are several questions about the validity of positivity and \ac{WGC} along the flow: what is the relation between their realization along a given \ac{RG} trajectory at finite $k$, and at its \ac{IR} endpoint, \ie{}, in the limit $k\to0$? Is it true that if positivity bounds are fulfilled/violated at the fixed point, they are also fulfilled/violated across the corresponding landscape?

While we will not try to prove any general statements, we will investigate the questions above for a sample of \ac{UV}-complete trajectories that flow into the \ac{GFP} in the \ac{IR}. To this end, we investigate a ``flowing'' version of Wilson coefficients, defined by removing the $k\to0$-limit in the original definitions, Eq.~\eqref{eq:WilsonCoeffDef-1} and~\eqref{eq:WilsonCoeffDef-2}, but still subtracting the logarithm along the whole flow as in~\eqref{eq:logsubtractioninWC}. We then insert these into the positivity bounds $\mathcal{P}_i$ and study them as a function of~$k$.

\begin{figure}
    \centering
    \includegraphics[width=0.55\textwidth]{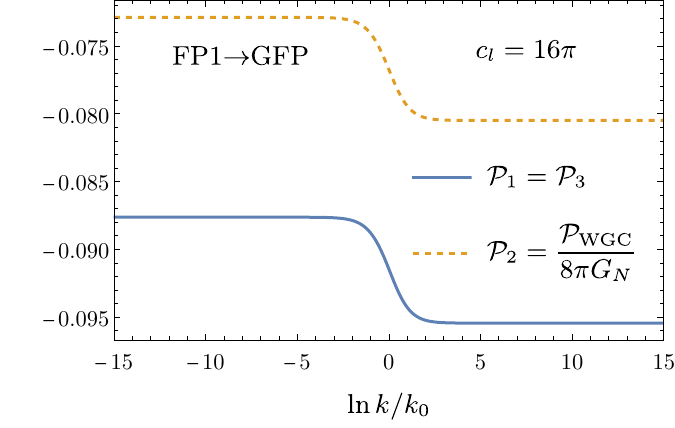}
    \caption{``Flowing'' positivity bounds along the separatrix from FP1 to the \ac{GFP}. We subtracted the logarithm in $g_-$ along the whole flow with $\scaleparameter{}=16\pi$. On this trajectory, $g_+>0$ and $g_{CFF}>0$, so that $\mathcal P_1=\mathcal P_3$ and $\mathcal P_2 = \mathcal P_{WGC}/8\pi G_N$. The two independent bounds $\mathcal P_1>0$ and $\mathcal P_2>0$ are violated along the entire flow, and the violation is an $\mathcal{O}(1)$ number in Planck units.}
    \label{fig:PosBoundFlow-separatrix}
\end{figure}

As a first example, we consider the flowing conditions for the separatrix between FP1 and the \ac{GFP}. For this trajectory, $g_+>0$ and $g_{CFF}>0$, so that $\mathcal P_1 \equiv \mathcal P_3$ and $\mathcal P_2\equiv\mathcal P_{WGC}/8\pi G_N$, and we are left with two independent bounds. As shown in \Cref{fig:PosBoundFlow-separatrix}, in this case the positivity bounds are (Planck-scale) violated along the entire flow.

\begin{figure}
    \centering
    \includegraphics[width=0.5\textwidth]{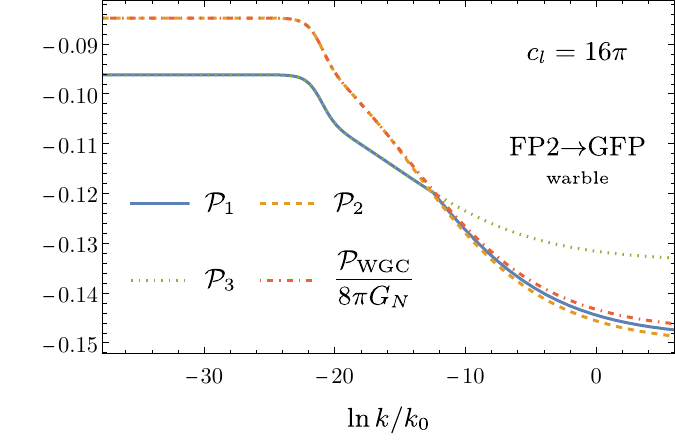}\hfill\includegraphics[width=0.5\textwidth]{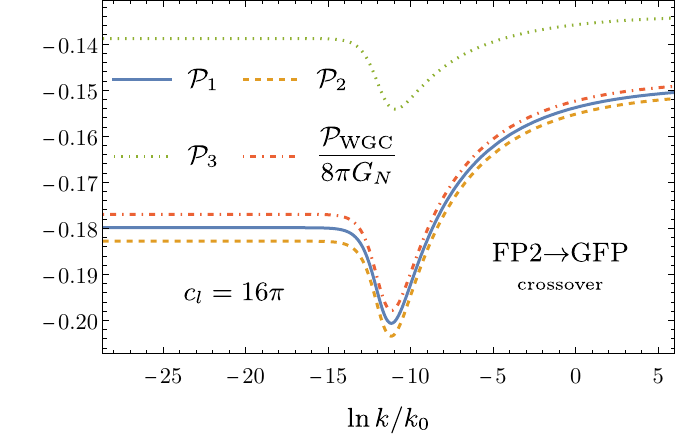}\\ ~ \\ \includegraphics[width=0.5\textwidth]{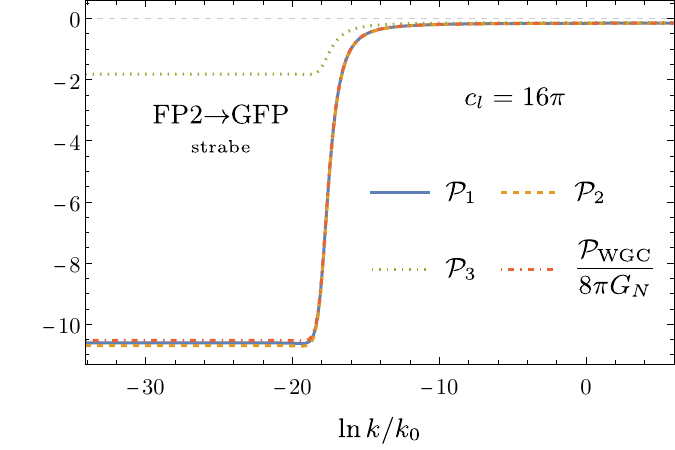}
    \caption{``Flowing'' positivity bounds for three different trajectories emanating from FP2 and reaching three different parts of the candy cane: close to the separatrix to FP1 (warble), close to the separatrix to \ac{MFP} (extended strabe), and in the crossover regime. We subtracted the logarithm in $g_-$ along the whole flow with $\scaleparameter{}=16\pi$. All trajectories display a violation of all positivity bounds along the entire flow. The violation becomes larger from the warble towards the extended strabe region, eventually surpassing an $\mathcal O(1)$ amount.}
    \label{fig:PosBoundFlow-FP2}
\end{figure}

A second set of examples can be extracted from the flow emanating from FP2. Trajectories originating at FP2 start out with a stronger violation of the flowing positivity bounds than the trajectory starting at FP1. The flow then behaves differently depending on its specific initial conditions: for trajectories ending up in the warble part of the landscape, the violation decreases, and approaches that of the FP1-\ac{GFP} separatrix. Near the crossover between warble and extended strabe, the violation stays approximately constant along the flow. Finally, in the extended strabe region, the violation increases along the flow. For these trajectories, $g_+$ and $g_{CFF}$ can be negative along the flow, so we do not necessarily have a degeneracy of the positivity bounds as for the separatrix between FP1 and the \ac{GFP}. In \Cref{fig:PosBoundFlow-FP2} we show these different behaviors.

In summary, in our system it is indeed true that the positivity bounds are Planck-scale violated along the whole flow for all trajectories that are \ac{UV}-complete and end up in the \ac{GFP} in the \ac{IR}. This behavior would support the idea that a relationship exists between the fulfillment/violation of positivity bounds at non-zero $k$ (including at the fixed point) and in the limit $k\to0$, where all quantum fluctuations are integrated out. While we refrain from drawing conclusions from these few examples, the above behavior once again emphasizes that FP1 comes with the least amount of violation of the bounds, even along the flow connecting it to the \ac{IR}.

\section{Summary and conclusions}\label{sec:conclusions}

In this work, we computed the landscape of \acp{EFT} stemming from \ac{UV}-complete (in particular, asymptotically safe) photon-graviton flows, and confronted it with entropy-based bounds that include the \ac{WGC} as a particular case~\cite{Cheung:2018cwt}, and, for the first time, with positivity bounds~\cite{Bellazzini:2019xts, CarrilloGonzalez:2023cbf}. The key idea is to generalize the notion of string landscape, which emerged within the swampland program~\cite{Vafa:2005ui, Palti:2019pca}, to other approaches to \ac{QG} and in particular to \ac{AS}~\cite{Basile:2021krr}. This has the ultimate scope of identifying the intersections of \ac{QG} landscapes, potentially highlighting connections between different approaches~\cite{deAlwis:2019aud,Basile:2021euh,Basile:2021krk}, systematically testing their consistency with bounds stemming from \ac{EFT}~\cite{deRham:2022hpx}, and assessing the validity of swampland conjectures beyond \ac{ST}, especially in the light of the String Lamppost Principle~\cite{Kumar:2009us,Montero:2020icj}.

As a first step in this endeavor, we focused on a gravitational system non-minimally coupled to an Abelian gauge field in four spacetime dimensions. Our dynamics includes (essential~\cite{Weinberg:1980gg}) operators up to the fourth order in a derivative expansion, totaling five interaction couplings. Within this setup, we found that the beta functions possess two gravitational fixed points that can serve as \ac{UV}-completion of the coupled theory, see \Cref{tab:FP}. The first has only one relevant direction, whose corresponding free parameter sets the scale of \ac{QG} and the unit with respect to which all other dimensionful quantities are measured. The ensuing sub-landscape is zero-dimensional, namely, it is a single point in the space of dimensionless Wilson coefficients. The second fixed point has two relevant parameters. Given that one of them sets the unit mass scale, its corresponding sub-landscape is a line in the shape of a stretched-out candy cane, see \Cref{fig:3DlandscapeFP2}. The tip of its small hook (the ``warble'') meets the sub-landscape point of the first fixed point, so that the two sub-landscapes are continuously connected. A large part of the whole landscape is nearly a straight line (the extended ``strabe''), whose other end is connected to a pure photon fixed point. Intriguingly, the entire landscape can be approximately embedded into a plane --- a non-trivial feature that has been observed before~\cite{Basile:2021krr}. Whether this is a coincidence or a universal feature of \ac{AS} is unclear, and deserves further systematic investigations.

We then confronted the set of Wilson coefficients within the \ac{AS} landscape with a collection of bounds stemming from different considerations, from positivity bounds motivated by unitarity and causality properties of scattering amplitudes~\cite{Bellazzini:2019xts, CarrilloGonzalez:2023cbf}, to a family of constraints based on the positivity of the entropy of black holes under the addition of higher order curvature operators. The latter also includes (one form of the) the electric \ac{WGC} as a particular case~\cite{Cheung:2018cwt}. We remark that positivity bounds are typically derived by excluding graviton fluctuations, and by assuming low spin dominance and a Regge limit~\cite{Herrero-Valea:2020wxz}. Violations of such bounds are generally expected upon the inclusion of gravitational fluctuations~\cite{Alberte:2020jsk, Alberte:2020bdz, Henriksson:2022oeu}, and are Planck-scale suppressed if appropriate combinations of dimensionless Wilson coefficients are of order one, cf. Eq.~\eqref{eq:planck-suppression}. Similar considerations also apply to the violation of the \ac{WGC}, where the allowed amount of negativity in the presence of a new physics scale was estimated in~\cite{Henriksson:2022oeu}. 

Consistent with these general expectations~\cite{Alberte:2020jsk, Alberte:2020bdz, Henriksson:2022oeu}, we found that both the positivity bounds of~\cite{Bellazzini:2019xts, CarrilloGonzalez:2023cbf} and the \ac{WGC} are violated across the entire \ac{AS} landscape, but that indeed for large parts of it, the violation is Planck-scale suppressed. In more detail, at the sub-landscape point of the most predictive fixed point, the violation is maximally suppressed. The amount of violation slightly increases along the warble part of the other sub-landscape. The further along the extended strabe one progresses, the larger the coefficients of the violation become. This behavior is not surprising, since in the asymptotic limit of the extended strabe, the pure photon (\ie{}, gravity-free) fixed point~\cite{Eichhorn:2021qet} is approached, where standard positivity bounds are violated.\footnote{A more thorough investigation, along the lines of~\cite{deBrito:2023myf,Gies:2024xzy}, is needed to understand the properties of this fixed point and its unitarity further.} We can also turn our perspective around: starting from this non-unitary self-interacting photon theory, Asymptotically Safe Gravity acts as a ``unitarizer'' since the addition of gravity and the requirement of maximizing predictivity bring the theory closer to fulfilling standard unitarity bounds.\footnote{This goes along with the idea that \ac{AS} provides an avenue to solve the triviality problem of the $U(1)$ sector~\cite{Christiansen:2017gtg}.} Indeed, as already highlighted, the most predictive theory (the single-point landscape) is the one minimizing the already Planck-scale-suppressed violations of positivity bounds, suggesting an ideal scenario that combines high predictivity with the eventual fulfillment of modified positivity bounds.

Along with positivity bounds, we also investigated a family of entropy-based positivity constraints, which include the \ac{WGC} as a particular case~\cite{Cheung:2018cwt}. Since these are off-shell bounds, they can (and do) include the coupling of the topological Euler term. Since the Gauss-Bonnet operator is a topological invariant, its coupling does not appear in any beta function, and thus generically does not have a fixed point~\cite{Falls:2020qhj, Knorr:2021slg}. As a consequence, the landscape-value of the Euler coupling is unconstrained, and within the current setup and state-of-the-art, such entropy-based bounds put constraints on the Euler coupling, rather than being a direct test of \ac{AS}. 

Our results are based on a setup that should be improved in future work. First, our computations were performed in Euclidean signature, making a Wick rotation necessary in order to relate the resulting \ac{AS} landscape to entropy and positivity bounds. Recent progress in the field~\cite{Fehre:2021eob} has shown that the spacetime signature does not strongly impact the flow, but it would be noteworthy to investigate whether this statement applies to the whole \ac{AS} landscape. Second, approximations had to be made, in our case by truncating the effective action to fourth order in derivatives. It is conceivable that improved approximations~\cite{Baldazzi:2023pep} and a better understanding of gauge and parameterization dependence of the flow~\cite{Gies:2015tca} could push the most predictive fixed point into a regime where standard positivity bounds are satisfied. Third, to achieve a more realistic description~\cite{Pastor-Gutierrez:2022nki}, an extended field content has to be included, which would contribute to the relevant Wilson coefficients. Last but not least, grounded on the discussion above, the role of the Euler coupling has to be clarified within \ac{AS}: it has to be understood what \ac{AS} can, or cannot say, about black hole entropy (along the lines of~\cite{Conroy:2015wfa, Platania:2023uda}) and, more generally, if a relation exists between bulk and boundary couplings, which could enhance the predictive power of off-shell quantities in this approach.

Finally, what would it mean if all these shortcomings were addressed, and \ac{AS} would still violate positivity bounds? There are different options in this case. The violation could simply indicate that \ac{AS} does not fulfill at least one of the assumptions underlying the derivation of positivity bounds. Potential candidates for these include the violation of microcausality or locality, which is not unrealistic for a theory of \ac{QG}. Alternatively, as mentioned earlier, the violation could be related to the low-spin dominance hypothesis not being realized, which is assumed for the derivation, but whose violation would not pose any obvious problems for \ac{QG}. Additionally, a violation could also mean that the mere existence of a \ac{UV} fixed point is not enough to obtain a unitary and causal theory. Indications for the latter were indeed found in the study of general non-perturbative scattering amplitudes~\cite{Draper:2020bop}. All these possibilities point to two key aspects within this endeavor: on the \ac{EFT} side, the necessity of understanding positivity bounds in the presence of gravitational fluctuations and beyond perturbative settings~\cite{Sinha:2020win,Guerrieri:2021ivu, Guerrieri:2022sod, Herrero-Valea:2022lfd}, with an improved understanding of the relation between violations and their sources. On the \ac{AS} side, the importance of mapping the \ac{UV} features of the \ac{AS} fixed point to the \ac{IR} properties of the corresponding landscape~\cite{Basile:2021krr}.

\acknowledgments

The authors would like to thank I.~Basile, A. Eichhorn, G.~N.~Remmen, C.~de~Rham, A.~Tokareva, and A.~Tolley for interesting discussions, and I.~Basile for feedback on the manuscript. The research of A.P. is supported by a research grant (VIL60819) from VILLUM FONDEN. A.P. also acknowledges support by Perimeter Institute for Theoretical Physics during the development of this project. B.K. is grateful for the hospitality of Perimeter Institute where part of this work was carried out. Research at Perimeter Institute is supported in part by the Government of Canada through the Department of Innovation, Science and Economic Development and by the Province of Ontario through the Ministry of Colleges and Universities. The research of B.K. is supported by Nordita. A.P. is grateful to Nordita for support within the “Nordita Distinguished Visitors” program and for hospitality during the early stages of this work. Nordita is supported in part by NordForsk.

\appendix

\section{Photon-graviton flows: setup and definitions} \label{sec:setup}

In this section, we review the framework of the \ac{FRG} and briefly discuss its modifications when accounting for field redefinitions. Afterward, we motivate our ansatz~\eqref{eq:action} for the effective action, and give details on the chosen gauge fixing and regularization.

\subsection{The Functional Renormalization Group}

 The \ac{FRG} is a powerful theoretical framework that is extensively employed in condensed matter physics, quantum field theory, and beyond~\cite{Dupuis:2020fhh}, as a prescription to evaluate the path integral and to investigate non-perturbative \ac{RG} flows. Unlike traditional \ac{RG} techniques, which focus on coarse-graining in momentum space, the \ac{FRG} operates directly in the space of actions, offering a more versatile approach to study strongly interacting systems. Specifically, the \ac{FRG} is based on a modification of the effective action by an \ac{IR} regulator. The latter has the dual scope of providing a prescription to regularize the path integral and of implementing the Wilsonian idea of shell-by-shell integration of modes. With the regularization implemented, the path integral can be translated into an exact functional differential equation~\cite{Wetterich:1992yh, Morris:1993qb, Ellwanger:1993mw} for the modified effective action~$\Gamma_k$,
\begin{equation}\label{eq:standardflowequation}
    k\partial_k \Gamma_k = \frac{1}{2} \mathrm{STr}\left\{\left(\Gamma_k^{(2)}+\regulatortensor_k\right)^{-1} \, k\partial_k \, \regulatortensor_k \right\} \, .
\end{equation}
Here, $\Gamma_k^{(2)}$ is the second functional derivative of $\Gamma_k$ with respect to the underlying fields, $\regulatortensor_k$ is the regulator implementing the successive mode integration, and STr denotes a so-called ``super-trace'', \ie{} a sum over discrete indices and an integration over continuous variables, together with a minus sign for Grassmann-valued fields. By construction, in the limit $k\to0$, we obtain the fully renormalized couplings and thus the corresponding Wilson coefficients. 

An essential ingredient for theories with gauge symmetries like gravity is the background field formalism. To be able to define a gauge fixing and a regularization, it is necessary to split the metric $g$ into an arbitrary background metric $\bar g$ and fluctuations $h$ about it,
\begin{equation}
    g_{\mu\nu} = \bar g_{\mu\nu} + h_{\mu\nu} \, .
\end{equation}
In our computation, we furthermore restrict ourselves to the so-called background field approximation, where we set $h$ to zero after having computed the two-point function $\Gamma_k^{(2)}$. For an overview of how to go beyond this, see \eg{}~\cite{Pawlowski:2020qer, Pawlowski:2023gym}.

Our focus in this work is to compute the \ac{AS} landscape, and to evaluate its intersections with \ac{IR} constraints like positivity bounds. To simplify the technical setup, we implement field redefinitions to eliminate as many inessential running couplings~\cite{Weinberg:1980gg} as possible. One minimal way of eliminating inessential couplings within the \ac{FRG} has been introduced in~\cite{Baldazzi:2021ydj}, and is based on an appropriate $k$-dependent field redefinition at the level of the flow equation,
\begin{equation}\label{eq:MESflow}
    k\partial_k \Gamma_k + \Psi_k \, \Gamma_k^{(1)} = \frac{1}{2} \mathrm{STr}\left\{\left(\Gamma_k^{(2)}+\regulatortensor_k\right)^{-1} \, \left[ k\partial_k + 2 \Psi_k^{(1)} \right] \, \regulatortensor_k \right\} \, .
\end{equation}
This equation is based on a more general flow equation~\cite{Wegner_1974, Pawlowski:2005xe}, see~\cite{Wetterich:1997bz, Gies:2001nw} for earlier applications. The modification resides in the \ac{RG} kernel $\Psi_k$, which is the expectation value of the flow of the redefined microscopic field. In practice, an ansatz is chosen for $\Psi_k$, and one has to check a posteriori whether this indeed corresponds to a well-defined field redefinition. Couplings that can be removed by~$\Psi_k$ are inessential, whereas all other couplings are the essential ones. The latter are also the only ones that can appear in observables like scattering amplitudes. The ``essential'' version of the \ac{FRG} flow has been applied widely, see \eg{}~\cite{Baldazzi:2021orb, Knorr:2022ilz, Ihssen:2023nqd, Knorr:2023usb, Baldazzi:2023pep}. A critical discussion is given in~\cite{Wetterich:2024uub}.

\subsection{Action, gauge fixing, regularization and field redefinitions}

The flow equations~\eqref{eq:standardflowequation} and~\eqref{eq:MESflow} can usually not be solved exactly, \ie{} for a generic $\Gamma_k$. The standard approach is to specify symmetry, field content, an ordering criterion for the operators, and a truncation order that is typically dictated by the computational limitations. The elimination of inessential couplings via the flow equation~\eqref{eq:MESflow} is a powerful instrument to push both truncation order and number of non-minimally coupled fields beyond previous computational limitations. In our work, we employ a derivative expansion of the effective action, and we include all essential couplings in \ac{QG} coupled to an Abelian gauge field with up to four derivatives. To this order, the essential part of the action reads
\begin{equation}\label{eq:Gamma}
\begin{aligned}
    \Gamma_k = &\int \text{d}^4x \, \sqrt{g} \, \Big[ \frac{1}{16\pi G_N} \left( 2\Lambda - R \right) + G_{\eulerterm} \, \eulerterm \Big] \\
    + &\int \text{d}^4x \, \sqrt{g} \, \Big[ \Ftwo{} + G_{\Ftwo^2} \, (\Ftwo{})^2 + G_{\Ffour} \, \Ffour{} + G_{CFF} \, C^{\mu\nu\rho\sigma} F_{\mu\nu} F_{\rho\sigma} \Big] \, ,
\end{aligned}
\end{equation}
where, as usual, all couplings depend on the scale $k$, and the operators $\Ftwo{}$, $\Ffour{}$, and $\eulerterm{}$ are those introduced in~\eqref{eq:operator-def}. Compared to the action in~\eqref{eq:action}, the above effective action includes a cosmological constant. Although the latter is the lowest-order operator in a derivative expansion, observationally $\Lambda G_N\ll1$, and thus, in a first approximation, we will consider $\Lambda=0$ in the limit $k\to0$. Based on these considerations, and following the steps in~\cite{Baldazzi:2021orb}, we use the field redefinition to fix~$\lambda$ to be proportional to~$g$, entailing a vanishing dimensionful cosmological constant for the effective action~\eqref{eq:action-intro}.

Since our system admits gauge symmetries, the effective action has to be complemented by a gauge fixing term. We employ the harmonic gauge in both sectors, so that
\begin{equation}
    \Gamma_\text{gf} = \int \text{d}^4x \, \sqrt{\bar g} \, \left[ \frac{1}{32\pi G_N} \left(\bar D^\alpha h_{\mu\alpha} - \frac{1}{2} \bar D_\mu h \right) \left(\bar D_\beta h^{\mu\beta} - \frac{1}{2} \bar D^\mu h \right) + \frac{1}{2} \left( \bar D^\mu a_\mu \right)^2 \right] \, .
\end{equation}
This gives rise to a Faddeev-Popov ghost action of the form
\begin{equation}
    \Gamma_\text{c} = \frac{1}{\sqrt{G_N}} \int \text{d}^4x \, \sqrt{\bar g} \, \bar c_{\mu} \left[ \bar \Delta \delta^\mu_{\phantom{\mu}\nu} - \bar R^\mu_{\phantom{\mu}\nu} \right] c^\nu + \int \text{d}^4x \, \sqrt{\bar g} \, \bar b \, \bar\Delta \, b \, ,
\end{equation}
where $\bar\Delta=-\bar D^2$. The next ingredient that we need to compute the \ac{RG} flow is the regulator. For this, we follow the argumentation of~\cite{Knorr:2022ilz} and introduce the ``natural'' endomorphisms in all sectors. This entails
\begin{align}
    \regulatortensor_k^h &= \frac{1}{32\pi G_N} \left[ \regulatorfunction(\bar \Delta_2) \Pi^\text{TL} - \regulatorfunction(\bar \Delta) \Pi^\text{Tr} \right] \, , \\
    \regulatortensor_k^a &= \regulatorfunction(\bar \Delta_a) \mathbbm 1 \, , \\
    \regulatortensor_k^c &= \regulatorfunction(\bar \Delta_c) \mathbbm 1 \, , \\
    \regulatortensor_k^b &= \regulatorfunction(\bar \Delta) \, .
\end{align}
We have identified all regulator shapes for simplicity, and $\Pi^\text{TL,Tr}$ denote the traceless and trace projectors, respectively. Eventually, we employed the Litim regulator~\cite{Litim:2001up},
\begin{equation}
    \regulatorfunction(x) = (k^2-x) \theta(1-x/k^2) \, .
\end{equation}
The operators used in the regulators read
\begin{equation}\label{eq:operators}
    \bar\Delta_{2\phantom{\mu\nu}\rho\sigma}^{\phantom{2}\mu\nu} = \left( \bar\Delta + \frac{2}{3} \bar R \right) \Pi^{\text{TL}\mu\nu}_{\phantom{\text{TL}\mu\nu}\rho\sigma} - 2 \bar C^{(\mu\phantom{\nu}\rho)}_{\phantom{(\mu}\nu\phantom{\rho)}\sigma} \, , \quad \bar\Delta_{a\phantom{\mu}\nu}^{\phantom{a}\mu} = \bar\Delta \delta^\mu_{\phantom{\mu}\nu} + \bar R^\mu_{\phantom{\mu}\nu} \, , \quad \bar\Delta_{c\phantom{\mu}\nu}^{\phantom{c}\mu} = \bar\Delta \delta^\mu_{\phantom{\mu}\nu} - \bar R^\mu_{\phantom{\mu}\nu} \, .
\end{equation}
The final ingredient for the \ac{RG} flow of essential couplings is the specification of the \ac{RG} kernel. Since we have two different fields, we can perform a field redefinition in the combined field space. At the order that we consider here, the most general corresponding \ac{RG} kernels are
\begin{equation}
\begin{aligned}
    \Psi^g_{\mu\nu} &= \gamma_g \, g_{\mu\nu} + \gamma_R \, R \, g_{\mu\nu} + \gamma_S \, S_{\mu\nu} + \gamma_{\Ftwo{}} \, \Ftwo{} \, g_{\mu\nu} + \gamma_{F^2}^\text{TL} \, \left( F_{\mu\alpha} F^\alpha_{\phantom{\alpha}\nu} + \Ftwo{} \, g_{\mu\nu} \right) \, , \\
    \Psi^a_{\mu} &= \gamma_a \, a_\mu + \gamma_{DF} \, D^\alpha F_{\mu\alpha} \, .
\end{aligned}
\end{equation}
Here, we set up the metric \ac{RG} kernel in a way to split it into trace and traceless parts, which disentangles the equations for the gamma functions maximally. To read off the beta and gamma functions, we complete the monomials in our action by the following invariants to form a basis:
\begin{equation}
    R^2 \, , \quad S^{\mu\nu} S_{\mu\nu} \, , \quad F^{\mu\nu} \Delta F_{\mu\nu} \, , \quad R \, F^{\mu\nu} F_{\mu\nu} \, , \quad S^{\mu\nu} F_{\mu\alpha} F^\alpha_{\phantom{\alpha}\nu} \, .
\end{equation}
This completes the discussion of the setup. With these ingredients as starting point, the computation of \ac{RG} flow has been performed with the help of the Mathematica package xAct~\cite{xActwebpage, Brizuela:2008ra, Nutma:2013zea} and a well-tested code~\cite{Knorr:2021lll, Knorr:2022ilz, Baldazzi:2023pep}. The complete set of beta functions can be found in the accompanying notebook~\cite{SupplementalMaterial}.

\section{Analytic Wilson coefficient in the pure matter theory} \label{sec:MFP_WC}

The analytic expression for the Wilson coefficient given in~\eqref{eq:MFP_WC} is
\begin{equation}\label{eq:MFP_exact_WC}
    \frac{G_{CFF}}{\sqrt{-G_+}} = -\sqrt{\frac{a_1}{\pi}} \frac{\Gamma(a_2)}{\Gamma\left(a_2+\frac{1}{2}\right)} \left[ {}_2F_1\left( \frac{1}{2}, a_3, a_2+\frac{1}{2} \bigg| z \right) + a_4 \, {}_2F_1\left( \frac{1}{2}, a_3+1, a_2+\frac{3}{2} \bigg| z \right) \right] \, ,
\end{equation}
where the numbers $a_{1,2,3,4}$ and $z$ are roots of low-order polynomials that can be obtained as follows. The number $a_1$ is the real root of the polynomial
\begin{equation}
\begin{aligned}
    10793861 + 18949368102912 x + 30552729884565700608 x^2 \\
    - 1619614712642678776922112 x^3 - 9100076856720841554681397248 x^4 \\
    - 16902672123436474482083424632832 x^5 + 3161447767348821628323748676370432 x^6
\end{aligned}
\end{equation}
near the point $a_1 \approx 0.0000177$. For $a_{2,3}$, we define the polynomial
\begin{equation}
\begin{aligned}
    413044310016 - 10879053926400 x \\
    + 158650842599040 x^2 - 1291967177365900 x^3 \\
    + 4894717105389125 x^4 - 8447086388113750 x^5 + 5813784153762500 x^6 \, .
\end{aligned}
\end{equation}
The number $a_2$ is then the real root of this polynomial near $a_2 \approx 0.113$, whereas $a_3$ is minus the root of the other real root of this polynomial, $a_3 \approx -0.313$. Next, the number $a_4$ is the real root of the polynomial
\begin{equation}
\begin{aligned}
    307642010808877056 + 179837560477922623488 x + 48915980589977271545856 x^2 \\
     + 5843687635359647850841088 x^3 + 826964630953265786764096 x^4 \\ + 39539603437261811380528 x^5 + 646308740102823594591 x^6
\end{aligned}
\end{equation}
near the point $a_4 \approx -0.00395$.
Finally, the number $z$ is the real root of the polynomial
\begin{equation}
\begin{aligned}
    5967 - 7673436 x + 3561700932 x^2 - 554287751986 x^3 \\
    + 3561700932 x^4 - 7673436 x^5 + 5967 x^6
\end{aligned}
\end{equation}
close to $z \approx 0.00303$.

\bibliographystyle{JHEP}
\bibliography{bib}

\end{document}